\pdfoutput=1

\documentclass[10pt,twocolumn]{article}

\usepackage[utf8]{inputenc}
\usepackage[T1]{fontenc}
\usepackage{mathptmx}
\usepackage[margin=1in,columnsep=0.25in]{geometry}
\usepackage{microtype}
\usepackage{graphicx}
\usepackage{booktabs}
\usepackage{array}
\usepackage{csquotes}
\MakeOuterQuote{"}
\usepackage[font=small,labelfont=bf,skip=5pt]{caption}
\usepackage{enumitem}
\setlist{itemsep=2pt,topsep=3pt,parsep=0pt}
\usepackage[colorlinks=true, linkcolor=blue, citecolor=blue, urlcolor=blue]{hyperref}
\hypersetup{
  pdftitle={The Commercial Tax: Rent-vs-Own Blind Spots in Multi-Hop Retrieval Benchmarks},
  pdfauthor={Luis M. Sanchez, Kosrow Dehnad},
  pdfkeywords={retrieval-augmented generation, multi-hop question answering,
               embedding models, software licensing, cost disclosure,
               benchmark reproducibility, enterprise AI adoption}}
\usepackage{xurl}
\usepackage{balance}
\usepackage{float}
\newcolumntype{L}[1]{>{\raggedright\arraybackslash}p{#1}}

\title{\textbf{The Commercial Tax:\\ Rent-vs-Own Blind Spots in Multi-Hop Retrieval Benchmarks}}
\author{
  Luis M. Sanchez\thanks{SGX Analytics} \\
  \texttt{luis.m.sanchez@toryx.ai} \and
  Kosrow Dehnad\thanks{Adjunct Professor, IEOR, Columbia University / Director, New York Institute and Laboratory for Artificial Intelligence} \\
  \texttt{kd11@columbia.edu}
}
\date{}

\newcommand{\ARTIFACTREPO}{\url{https://github.com/Toryx-AI/commercial-tax-multihop-retrieval}}
\newcommand{\ARTIFACTDOI}{\href{https://doi.org/10.5281/zenodo.21972866}{10.5281/zenodo.21972866}}
\newcommand{\ARTIFACTHF}{\href{https://huggingface.co/datasets/toryx-ai/commercial-tax-musique-embeddings/tree/bd00a9c32c00fe9007f9244475f1b11ba7d51f69}{\texttt{toryx-ai/commercial-tax-musique-embeddings}@\texttt{bd00a9c3}}}
\newcommand{\ARTIFACTDATE}{2026-08-17}

\begin{document}

\twocolumn[
\maketitle
\begin{abstract}
Enterprises connect language models to their own data through retrieval. The benchmarks that rank multi-hop retrieval systems leave out two facts a buyer needs before a published number can be used: whether the retrieval backbone may be deployed commercially, and what it costs to build. On licensing: the field's dense-retrieval anchor, NV-Embed-v2, is licensed \texttt{cc-by-nc-4.0}, non-commercial. Of the four leading MuSiQue systems we audit (HippoRAG-2, PropRAG, SAG, KET-RAG), three depend on it for their best numbers and none says so; the fourth builds on a commercial API embedder. On performance: we measure thirteen embedders from eight makers on one identical MuSiQue harness, with bootstrap confidence intervals throughout. Until mid-2026 there was a real commercial tax: the best commercially-licensed embedder in this panel trailed the anchor by 2.31 Recall@5 points (95\% CI {[}0.91, 3.71{]}, $p=0.001$; the pre-specified primary comparison, reported uncorrected). NVIDIA's Nemotron-3-Embed-8B, released 2026-07-16, has closed it: $+0.24$ at Recall@5 (95\% CI {[}$-0.94$, $+1.43${]}, $p=0.69$) and $-0.58$ at Recall@10 (95\% CI {[}$-1.65$, $+0.48${]}, $p=0.28$). It matches the anchor; it does not beat it, and at $n=1{,}000$ the design could not have detected a difference smaller than about 1.7 points. It is the only entrant that is commercially licensed, free to self-host, \emph{and} indistinguishable from the anchor. Every other entrant meeting the first two conditions sits 5.2 to 14.6 points below the anchor.\footnote{Five entrants meet both conditions: Nemotron-3-Embed-1B ($-5.23$), Llama-Nemotron-Embed-1B-v2 ($-5.82$), Qwen3-VL-Embedding-8B ($-9.67$), mxbai-embed-large-v1 ($-13.84$) and BGE-M3 ($-14.62$), all against the anchor at Recall@5 (\texttt{panel\_anchor\_family.json}). \texttt{nv-embedqa-e5-v5} is commercially licensed but reachable only as a NIM platform service, not as weights, so it is not in this set; its $-11.86$ falls inside the range regardless. The API-only entrants are excluded by the self-hosting condition, and one of them, voyage-3.5 at $-15.47$, sits further below the anchor than any model in this set.} The durable finding is not the ranking but the paid-versus-free divide: API embedders charge per token on every re-index, self-hosted ones charge nothing. On cost: three of the five systems in our cost audit (the four above plus Microsoft's GraphRAG) do not disclose what indexing costs. Two report no dollar figure at all; a third reports token counts but no dollars. The two that do publish a figure price a 5.64\,MB academic corpus, and the only published GraphRAG dollar figures span 11x inside one third-party paper (\$2.30 for a low-cost configuration, \$24.94 for a high-performance one, to index that corpus once); an independent token-based reconstruction (\$19.49--\$38.99) brackets the latter. Extrapolated linearly to 1\,TB, that undisclosed configuration choice separates roughly \$428K from roughly \$4.6M to index once. Our cost model keeps one-time embedding cost apart from recurring answering cost, because a one-time build and an annual sum do not belong under one multiplier. At a matched 1\,TB, embedding sits 7.5x to 900x below graph construction, and a full year of answering at 10,000 queries a day sits 350x or more below it.

\vspace{0.5em}
\end{abstract}
\vspace{1.5em}
]

\section{Introduction}
\label{sec:intro}

Retrieval-augmented generation is how industry actually connects large language models to proprietary data, and by a wide margin. The peer-reviewed evidence is qualitative but it points one way: practitioners interviewed across industry prefer retrieval over fine-tuning because a knowledge base "can be easily updated and expanded" and because "the system must present the sources of information transparently" \cite{brehme2025}. Quantitative adoption figures exist, but they come from vendors with a commercial interest in the answer, so we cite them as indicative and no more. A vendor survey puts augmented-LLM use at 86\% of organizations against 14\% on generic models \cite{k2view2024}. Databricks reports vector-database usage on its own platform growing 377\% year over year \cite{databricks2024}. We rely on the direction these agree on, not their digits. The retrieval layer is not a niche academic concern. It is the load-bearing wall of enterprise AI adoption.

\begin{figure*}[t]
\centering
\includegraphics[width=0.85\textwidth]{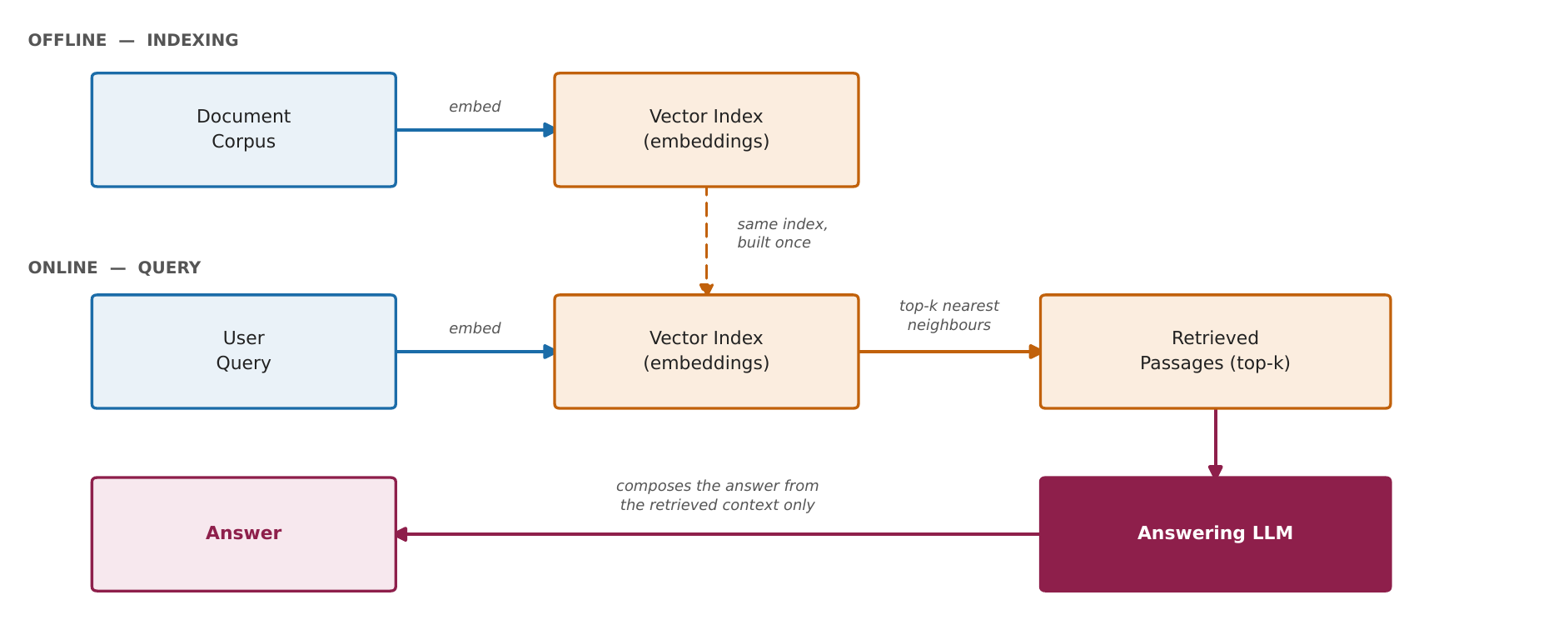}
\caption{Retrieval-augmented generation: offline indexing builds a vector index once from the document corpus; online, a query is embedded, matched against that same index for top-$k$ nearest neighbors, and the retrieved passages are handed to an LLM to compose an answer.}
\label{fig:rag-architecture}
\end{figure*}

Against that backdrop, the benchmarks the research community uses to measure multi-hop retrieval progress report numbers that look transferable to industry and, on inspection, are not. The chief one is MuSiQue \cite{trivedi2022musique} under the closed-corpus protocol standardized by HippoRAG-2 \cite{gutierrez2025}. To make "multi-hop" concrete, take one question from the 1,000-question set our harness runs: \textit{"What is the name of the famous bridge in the birth city of the composer of Scanderbeg?"} It can only be answered by chaining three facts. Scanderbeg's composer is Vivaldi; Vivaldi was born in Venice; Venice's famous bridge is the Rialto. No single passage holds the answer. The retriever has to surface all three linked passages, and that is why Recall@$k$ is the metric throughout this paper: the percentage of the source passages that were needed and that appear in the top $k$ search results. The subject matter is Wikipedia's. The structure is not, and the structure is what transfers. The same three-link shape in an institutional corpus reads \textit{"which reinsurer wrote the treaty covering the terminal operated by the counterparty named in this credit agreement?"} The counterparty sits in the agreement, the terminal in an operating filing, the treaty in a third document. No one of them answers the question. A retriever that surfaces two of the three returns an answer that is confidently wrong. That is the failure mode this metric exists to catch.

Two other terms recur and are worth fixing here for readers who are not IR specialists.

A \emph{dense-retrieval anchor} is the baseline search model a benchmark's published results are measured against. "Standard" here means standard \emph{for this protocol and everyone who reports into it}: HippoRAG-2's authors fixed the anchor when they defined the evaluation, and each subsequent system (PropRAG, SAG, and the reproductions that cite them) kept it so their numbers would stay comparable, with the result that an entire comparison table inherits whatever properties that one component has. The anchor under audit is NV-Embed-v2.

A \emph{percentile bootstrap} interval is a confidence interval computed by re-drawing the evaluation questions with replacement many times (100,000 times here for the paper's headline tests), then reading off the middle 95\% of the resulting scores. It is what lets us say whether a gap between two embedders is real or is within the noise of which questions happened to be asked.

This paper's central claim is simple. These benchmark numbers are reported stripped of the two variables that decide whether they mean anything to a buyer: what you are legally allowed to deploy, and what it costs to build. Everything that follows measures one or the other.

\paragraph{Licensing.} As of mid-2026, every leading system in this space anchors on the same dense-retrieval backbone, NV-Embed-v2, and NV-Embed-v2 is licensed \texttt{cc-by-nc-4.0}: non-commercial use only. NVIDIA's own model card is explicit. \textit{"This model should not be used for any commercial purpose."} We verify this from primary sources and trace it to its root cause (\S\ref{sec:background-license}). We then audit the three leading systems that rely on it, HippoRAG-2~\cite{gutierrez2025} and PropRAG~\cite{wang2025proprag} as their embedding backbone and SAG~\cite{wu2026sag} in its best-reported ablation, and find that none discloses the restriction. The fourth system we audit, KET-RAG~\cite{huang2025ketrag}, does not use the anchor at all; it builds on OpenAI's \texttt{text-embedding-3-small}. That makes it the counterexample, a leading system showing that a commercially-usable embedder was a live option. A practitioner reading any of these papers has no way to know that the number in front of them is not commercially deployable as published.

\paragraph{Cost.} The retrieval structures these systems build (knowledge graphs, proposition indices, SQL schemas) require running an LLM over the entire corpus, at a cost that scales with corpus size. Two systems disclose a dollar figure for the shared corpus: PropRAG about \$4, KET-RAG \$1.89. The protocol paper, HippoRAG-2, discloses token counts from which a dollar cost must be derived. Others discuss cost as a design concern and report no figure at all; SAG is one, and so is Microsoft's own GraphRAG paper~\cite{edge2024graphrag}. The \$33K figure widely associated with GraphRAG is a third-party extrapolation from KET-RAG. It is not Microsoft's number. A reader cannot answer "what would this cost my organization" from any of these papers' own text.

Neither gap is exotic. Both are the same gap: a headline number reported without the conditions under which it would be deployed. A published number that omits those conditions is not a wrong number; it is a number measured under conditions the reader does not share, and quantitative finance learned to correct for exactly this decades ago. The authors of this paper come to it from quantitative finance, where that failure mode has an exact and well-studied name: the strategy that looks brilliant in a backtest and underperforms in production because the backtest silently left out transaction costs. Slippage, commissions, market impact, the borrow on a short. Every desk has a story about a signal that was real on paper and gone the moment it had to pay to trade. No serious desk evaluates a strategy on gross paper returns; net of cost is the only number that counts, and a strategy that will not survive its own frictions is not a strategy. This paper is that discipline transplanted. Measure the headline quality number \emph{and} its deployment frictions on the same harness, and report both.

\paragraph{Non-claims.} We do not claim that NV-Embed-v2-anchored research is invalid, that the field's evaluation protocols are unsound, or that any paper engaged in misconduct. We claim that the standard MuSiQue/HippoRAG-2 protocol anchors on a component that is non-commercially licensed and cost-undisclosed, and that this leaves the question a corporate buyer actually has unanswered by the published record: what may we legally deploy, and what will it cost us to run. Both gaps can be measured. This paper measures them.

\paragraph{Research questions.}
\begin{enumerate}[leftmargin=1.8em]
\item License verification (\S\ref{sec:task-a}/\S\ref{sec:res-license}): What are the exact, current licenses of the embedding models used in MuSiQue benchmarking?
\item Commercial-embedder retrieval floor (\S\ref{sec:task-b}/\S\ref{sec:res-panel}): What is the retrieval-quality gap between the non-commercial anchor and a vendor-neutral panel of commercially-licensed alternatives, measured identically?
\item Literature disclosure audit (\S\ref{sec:task-cd}/\S\ref{sec:res-audit}): Do published systems disclose the license and cost of the components their headline numbers depend on?
\item Standardized cost model (\S\ref{sec:cost-model}/\S\ref{sec:res-index-cost}--\ref{sec:res-f1}): What does indexing cost, and what does answering a query cost, at realistic institutional scale?
\end{enumerate}

\section{Background}
\label{sec:background}

\subsection{MuSiQue and the HippoRAG-2 Protocol}
\label{sec:background-musique}

MuSiQue \cite{trivedi2022musique}, short for \emph{Multihop Questions via Single-hop Question Composition}, was built to fix a known weakness of HotpotQA. HotpotQA \cite{yang2018hotpotqa} is the earlier and still widely-cited multi-hop benchmark: 113,000 questions written by crowdworkers over pairs of Wikipedia articles, each labeled with the sentences a human used to answer it. Its weakness is that many of those crowdsourced 2-hop questions turned out to be largely solvable without genuine multi-hop reasoning: a model could shortcut to the answer from one passage, or from surface cues, without ever traversing the chain. MuSiQue composes multi-hop questions bottom-up from chains of single-hop questions and filters out shortcut-answerable compositions with trained-model disconnection probes, making it, as of early 2026, the hardest widely-used closed-corpus multi-hop benchmark. The HippoRAG-2 protocol \cite{gutierrez2025} standardized evaluation on a fixed pool of 11,656 Wikipedia passages and a 1,000-question sample of MuSiQue's dev set, which its authors describe as randomly collected and release as a file.\footnote{We take that file as distributed (\texttt{musique.json}, 1,000 questions, from the HippoRAG-2 release) rather than re-sampling 1,000 questions ourselves, which is what makes our 69.55 directly comparable to the 69.7 its own table reports; the released question order is the harness's order, and the per-question vectors we release are aligned to it index-by-index. Nothing in this paper depends on which 1,000 MuSiQue questions the protocol's authors drew, only on every entrant answering the same ones.} It fixes the dense-retrieval floor at NV-Embed-v2 (69.7\% Recall@5 in HippoRAG-2's own result table) so that graph-augmented (HippoRAG-2~\cite{gutierrez2025}, 74.7\%), proposition-guided (PropRAG~\cite{wang2025proprag}, 78.3\%), and SQL-structured (SAG~\cite{wu2026sag}, 80.0\%) systems compare apples-to-apples. SAG is worth a second sentence here: its headline 80.0\% uses BGE-Large-EN-v1.5, but its own ablation reports 81.7\% (81.71\% in the authors' released benchmark repository) when the embedder is swapped to NV-Embed-v2. The highest MuSiQue Recall@5 in this group is produced by the non-commercially-licensed anchor, in a table whose stated purpose is to show the system does not depend on it. The protocol's internal rigor is genuine; our critique concerns what it leaves undisclosed about the anchor component underneath it. The five audited systems span academic and industry labs: Ohio State, UIUC, Zleap AI, NUS, and Microsoft Research.

We independently measured the 11,656-passage corpus at 5.64MB of raw text, averaging 79.8 words per passage. That number is load-bearing for the cost extrapolations of \S\ref{sec:cost-model}.

\subsection{Why NV-Embed-v2 Specifically Is Non-Commercial}
\label{sec:background-license}

NV-Embed-v2 \cite{lee2024} is the second-generation model in NVIDIA's NV-Embed line and the dense-retrieval floor of the protocol in \S\ref{sec:background-musique}. Permissive licenses (Apache 2.0, MIT) permit commercial deployment; CC-BY-NC-4.0~\cite{cc2013} prohibits it outright. NV-Embed-v2's card~\cite{nvidia2024nvembed} states the restriction without ambiguity and directs commercial users to NVIDIA's separately-licensed, paid NeMo Retriever Microservices (NIMs). That is a real commercial path, not an academic dead end.

The restriction is traceable, not incidental: NV-Embed's training mix includes MS MARCO, Microsoft's benchmark corpus, whose license is "intended for non-commercial research purposes only." An NVIDIA representative stated the general mechanism publicly, on the model's own HuggingFace discussion board:\footnote{The comment was posted on 9 May 2025 by the account \texttt{nada5}, badged as a member of the \texttt{nvidia} organization on HuggingFace, in discussion \#40 on the \texttt{nvidia/NV-Embed-v2} model repository, the same repository that hosts the weights and the model card. We treat it as an authoritative statement of the licensing rationale because of that organizational badge and its location on NVIDIA's own model page, not as a formal corporate position; NVIDIA has published no separate statement on the question. The thread remains publicly readable at the URL in \cite{nvidia2025discussion}.} \textit{"NV-Embed is a research model which employed the non-commercial datasets for training, so we could not declare it's commercial model"} \cite{nvidia2025discussion}. That comment names non-commercial training data as the cause but does not name a specific dataset; MS MARCO is our identification, from the training mix disclosed in NV-Embed's own paper. The CC-BY-NC-4.0 license is, in effect, NV-Embed-v2 inheriting the most restrictive term in its own training data, and the restriction is \emph{not} defeated by accessing the model through NVIDIA's NIM API rather than downloading the weights. NIM (NVIDIA Inference Microservices) is NVIDIA's packaging of a model as a hosted, containerised inference endpoint: the caller sends text to an API and receives embeddings, with NVIDIA, not the caller, running the GPUs. It changes who operates the hardware, not what the model is licensed for: a different access channel to the same restricted model, not a separate license.

Longpre et al.~\cite{longpre2024} audited 1,800+ AI \emph{datasets} and found over 70\% had missing or incorrect license disclosure. We apply the same discipline one layer down, at the model-dependency level, and, separately, to cost disclosure, which to our knowledge has not been systematically audited in this literature before.

\section{Methodology}
\label{sec:method}

\paragraph{Metrics.} An \emph{embedder} maps text to a fixed-length vector such that semantically similar texts receive similar vectors; retrieval embeds the corpus once offline, embeds each query online, and returns the top-$k$ passages by cosine similarity (Figure~\ref{fig:rag-architecture}; a visual explainer is in Appendix~\ref{app:worked}). \emph{Recall@$k$} is the fraction of a question's gold supporting passages appearing in the retrieved top-$k$. A fully worked example is in Appendix~\ref{app:worked}.

\subsection{Task A: License Verification}
\label{sec:task-a}

We verified licenses for NV-Embed-v2 and every panel alternative directly from HuggingFace model-card YAML frontmatter and card text (and, for API-only providers, the provider's own current documentation), timestamped per-model at first verification. A model passes if its license permits commercial deployment without a separate paid license or non-commercial clause.

\subsection{Task B: Commercial-Embedder Retrieval Floor}
\label{sec:task-b}

This measurement introduces no architecture of our own: embed each of the 1,000 evaluation questions with no decomposition, rewriting, or reranking; embed the 11,656-passage corpus once; retrieve top-$k$ by cosine similarity; score Recall@$k$ against gold passages. It is the identical "dense retrieval alone" floor test the literature reports for NV-Embed-v2. We ran it ourselves, on one harness, for all thirteen embedders, \emph{including NV-Embed-v2 itself}.

The panel spans eight makers: NVIDIA (NV-Embed-v2 research anchor; Nemotron-3-Embed-8B and -1B; Llama-Nemotron-Embed-1B-v2; \texttt{nv-embedqa-e5-v5}), OpenAI (text-embedding-3-small/-large), Cohere (Embed v4 via Amazon Bedrock), Google (Google Gemini embedding-001), Voyage AI (voyage-3.5), and three self-hosted open-weight models (Qwen3-VL-Embedding-8B, mxbai-embed-large-v1, BGE-M3). Five of the thirteen are NVIDIA models, so the panel is not vendor-neutral, and we say why. The research anchor under audit is NVIDIA's; the licensing question this paper asks is specifically about NVIDIA's decision to ship that anchor non-commercially while shipping other models commercially; and answering it requires testing the commercial line NVIDIA directs those users to. The winning model therefore shares a maker with the anchor it matches, and readers should weigh the headline accordingly. The competitive context is supplied by the other eight entrants, none of them NVIDIA's. Full composition, licenses, and access channels are in Appendix~\ref{app:panel} (Table~\ref{tab:panel}). Two formatting accommodations are applied, both standard practice. For each open-weight embedder we tested several query-formatting variants and kept the best. For each API provider we used the provider's own documented \emph{asymmetric query/document mode}, meaning that questions and documents are embedded differently, on purpose. The provider embeds a short question and a long passage through different code paths: the caller declares which role the text plays (Cohere's \texttt{input\_type}, Gemini's \texttt{task\_type}) and the provider applies a different instruction or projection to each. Ignoring it embeds a question as though it were a document, which measurably degrades retrieval. Skipping either accommodation would unfairly penalize the affected model (format-sensitivity findings in \S\ref{sec:res-panel}).

Keeping the best of several variants is itself a choice, and it flatters whichever model got the most attempts, by an amount the confidence interval does not capture, because that interval is computed for the winning variant alone. So we report how many attempts each entrant received; a reader should not have to assume the number was uniform. It was not. NV-Embed-v2, the anchor, was swept over four query instructions on the matched corpus ($69.42$, $69.53$, $\mathbf{69.55}$, $66.90$; the bolded best is the \texttt{web-search} instruction, released as \texttt{web-search|title+text} in the per-question vectors, and it is that variant which is carried forward into every table and figure here. The same four instructions on the text-only index are reported in Appendix~\ref{app:reproduction}, where the best is again \texttt{web-search} at 67.09). Qwen3-VL was swept over a range spanning $11.6$ points. Nemotron-3-Embed-8B was measured in a single configuration, as was every API entrant.

That unevenness in attempts runs \emph{against} our own headline, not for it. Nemotron-3-Embed-8B's $+0.24$ lead is one configuration measured against the anchor's best of four. That makes it a conservative estimate of the gap. Score the anchor on the mean of its sweep instead of its maximum and the gap \emph{widens} to $+0.29$ (the three instructed variants: 69.42, 69.53, 69.55) or $+0.94$ (all four, adding the uninstructed 66.90), both inside the interval, though that interval was computed for the best-variant pairing and is quoted here for scale only. The entrant most advantaged by our policy is the non-commercial anchor whose position this paper argues against. Readers should weigh the open-weight rows accordingly.\footnote{Nemotron-3-Embed-8B's $+0.10$ run-to-run $\Delta$ in Table~\ref{tab:reproduction} is reproducibility of that same single configuration, not a second variant. Every figure reported here, the anchor's included, comes from one re-measurement campaign; substituting Nemotron's earlier run would still leave the headline a match, at $+0.14$.}

\subsection{Tasks C and D: Literature Disclosure Audits}
\label{sec:task-cd}

We audit four leading MuSiQue-benchmarking papers (HippoRAG-2, PropRAG, SAG, KET-RAG) for whether they disclose the license of their retrieval-backbone embedder, searching each paper's full text and linked code repositories for the terms \emph{license}, \emph{licence}, \emph{commercial}, \emph{non-commercial} and \emph{CC-BY} (both spellings, since the corpus is not uniformly US-spelled). A later pass over the four papers' full text added the license strings themselves, \emph{cc-by-nc}, \emph{nc-4.0} and \emph{noncommercial}; none appears in any of the four. For KET-RAG the question does not arise: its backbone is OpenAI's commercial API embedder rather than the anchor. Stating the search terms matters because the finding is a negative one: a claim that something is absent is only as strong as the search that failed to find it, and a reader should be able to re-run ours. For SAG we additionally cloned the linked benchmark repository and searched every tracked file, including both README variants, the license file, and the packaging and environment manifests; the only license text found anywhere was the MIT license covering the authors' own code. In parallel (adding Microsoft's GraphRAG paper, which the field treats as a baseline), we audit whether each system's \emph{own} paper discloses a dollar cost for indexing, fetching each paper's arXiv source directly and searching for currency markers instead of relying on summaries. That search is stated for the same reason the license one is, since it also reports an absence: the terms are \texttt{\$}, \emph{USD}, \emph{dollar}, \emph{cost}, \emph{price}, \emph{spend} and \emph{budget}, over each paper's full text including tables and appendices.

\paragraph{How the audited set was chosen.} It is a relevance-seeded snowball sample, not a systematic review, and the distinction governs how its counts should be read. KET-RAG was selected first, because it prices Graph-RAG indexing directly and so speaks to the cost question this paper asks; the other systems were reached from its related work and from the protocol paper's citation neighbourhood, keeping those published in 2025--2026 that report MuSiQue retrieval under HippoRAG-2's closed-corpus protocol. Microsoft's GraphRAG was added to the cost audit only, as the baseline the field prices against. Where we call three of them \emph{leading} the ground is specific and checkable: they hold the highest published Recall@5 under this protocol (74.7\%, 78.3\%, and 80.0\%/81.7\%). It is not an impression of prominence. What follows from the sampling is that every count in this paper is a count over the audited set and never a prevalence claim about the field: ``none of the three'' and ``two of five'' describe these papers. Establishing a rate for the literature would need a systematic search we did not run. What this set can establish is that the omission occurs at the top of the leaderboard, which is where a practitioner copying a number is most likely to be reading.

\subsection{Standardized Cost Model: Two Axes, Kept Separate}
\label{sec:cost-model}

Because two audited systems disclose no indexing cost, and the two that do report figures on a 5.64MB academic corpus far below deployment size, we build a transparent, rate-based model with two axes that behave differently and must not be conflated:

\begin{enumerate}[leftmargin=1.8em]
\item \emph{Embedding cost: one-time, scales with corpus size.} Each API provider's real published per-token rate applied to our measured corpus ($\sim$1.27M tokens; the $\sim$1.6\% that are evaluation queries are scaled with the corpus for simplicity, overstating extrapolated figures by at most that fraction) and extrapolated to a 10GB illustrative corpus ($\sim$1,816x our measured 5.64MB). Every self-hostable open-weight model costs \$0 per token in provider charges at any scale, at any volume; the tradeoff is a one-time hardware cost, which we anchor at a single NVIDIA DGX Spark~\cite{nvidia2026spark} (\$4,699, July 2026), deliberately priced as the whole-pipeline appliance covering both embedding and the heavier LLM workloads of the planned follow-on, not as an embedding-only requirement (rationale in Appendix~\ref{app:costnotes}).
\item \emph{Answering cost: recurring, scales with query volume.} gpt-4o-mini at published rates (batch \$0.075/1M input, \$0.30/1M output) as a fixed LLM over the real top-$k$ retrieved passages, for all 1,000 questions, at both k=5 and k=10, measured from real API usage, not estimated. Decoding is deterministic (\texttt{temperature=0.0}, \texttt{max\_tokens=32}, single sample per question); the exact system prompt, user-prompt template and passage serialisation are given in Appendix~\ref{app:costnotes}. This cost is independent of corpus size (the LLM only ever sees $k$ passages); it scales with how many questions get asked.
\end{enumerate}

\paragraph{The line that is absent, and why.} A third cost line appears in most production quotations and in none of the figures here: the vector store. Retrieval in this paper is exact brute-force cosine over all 11,656 passage vectors: every query is scored against every passage, with no approximate-nearest-neighbour index anywhere in the pipeline. That is a deliberate methodological choice before it is a cost one: an ANN index trades recall for speed, and any such loss would be indistinguishable, in the final Recall@$k$, from a weakness in the embedder under test. Exhaustive search removes that confound, so every point of difference reported here belongs to the embedding model and not to the index that stores it. At institutional scale the same vectors are served from a self-hosted vector database (Qdrant in our case; any comparable engine would do) on local NVMe: a fixed hardware cost, already inside the appliance price above, carrying no per-query fee and no per-gigabyte-per-month storage rent. A managed vector service would add a recurring line to Table~\ref{tab:indexing-cost} that self-hosting simply does not have. Its absence from our cost model is not an omission; it is the same rent-versus-own distinction the license axis makes, applied to storage. One consequence of that production path: it is \emph{approximate}. Those collections are created with the engine's default approximate index (HNSW, a graph-based nearest-neighbour structure that trades a little recall for speed) and queried with no exactness override, so deployment recall carries an ANN loss this paper does not measure. Every Recall@$k$ reported here is thus the index-free ceiling for its embedder. That is the right quantity for comparing embedders, since none of them is indexed and the confound is removed for all thirteen alike. It is not a production figure. A buyer should read 69.55\% as what the model can retrieve, then ask separately what their index gives back.

We fix a single LLM and do not compare several: the LLM-tier cost/accuracy tradeoff is left to the planned follow-on (\S\ref{sec:conclusion}).

\section{Results}
\label{sec:results}

\subsection{License Verification}
\label{sec:res-license}

NV-Embed-v2 is \texttt{cc-by-nc-4.0} ("should not be used for any commercial purpose"), traced in \S\ref{sec:background-license} to MS MARCO's non-commercial terms. Both Nemotron-3-Embed tiers, Llama-Nemotron-Embed-1B-v2, and \texttt{nv-embedqa-e5-v5} are commercial (OpenMDW-1.1, NVIDIA Open Model License, and AI Foundation Models Community License + MIT respectively), each trained without MS MARCO. mxbai-embed-large-v1 and Qwen3-VL-Embedding-8B are \texttt{apache-2.0}; BGE-M3 is \texttt{mit}. The OpenAI, Cohere, Google, and Voyage embedding APIs are standard commercial products under their terms of service.

\subsection{The Commercial-Embedder Retrieval Floor}
\label{sec:res-panel}

\begin{table*}[t]
\centering
\footnotesize
\setlength{\tabcolsep}{4pt}
\begin{tabular}{@{}llccccc@{}}
\toprule
Embedder & Maker & License & Recall@5 & 95\% CI & Recall@10 & 95\% CI \\
\midrule
Nemotron-3-Embed-8B (self-host) & NVIDIA & Commercial & 69.79 & [68.13, 71.43] & 77.54 & [75.96, 79.09] \\
NV-Embed-v2 (research anchor) & NVIDIA & Non-commercial & 69.55 & [67.89, 71.20] & 78.12 & [76.58, 79.63] \\
Gemini embedding-001 & Google & Commercial & 67.24 & [65.45, 69.01] & 76.35 & [74.71, 77.97] \\
Nemotron-3-Embed-1B (NIM) & NVIDIA & Commercial & 64.32 & [62.63, 65.98] & 72.79 & [71.13, 74.43] \\
Llama-Nemotron-Embed-1B-v2 (NIM) & NVIDIA & Commercial & 63.73 & [62.03, 65.42] & 71.35 & [69.69, 73.01] \\
Cohere Embed v4 (Bedrock) & Cohere & Commercial & 60.21 & [58.45, 61.98] & 69.08 & [67.34, 70.81] \\
Qwen3-VL-Embedding-8B & Alibaba & Free/open (\texttt{apache-2.0}) & 59.88 & [58.16, 61.60] & 68.53 & [66.88, 70.18] \\
OpenAI text-embedding-3-large & OpenAI & Commercial & 59.48 & [57.68, 61.29] & 70.15 & [68.41, 71.89] \\
nv-embedqa-e5-v5 (NIM) & NVIDIA & Commercial & 57.69 & [55.98, 59.40] & 66.27 & [64.53, 67.99] \\
mxbai-embed-large-v1 & Mixedbread AI & Free/open (\texttt{apache-2.0}) & 55.71 & [53.85, 57.56] & 64.35 & [62.50, 66.19] \\
OpenAI text-embedding-3-small & OpenAI & Commercial & 55.38 & [53.62, 57.12] & 64.78 & [63.04, 66.49] \\
BGE-M3 & BAAI & Free/open (\texttt{mit}) & 54.93 & [53.23, 56.62] & 62.89 & [61.17, 64.61] \\
Voyage voyage-3.5 & Voyage AI & Commercial & 54.08 & [52.24, 55.93] & 63.68 & [61.85, 65.49] \\
\bottomrule
\end{tabular}
\caption{Recall@5/@10 across the full 13-embedder panel, one identical harness, every passage embedded as \texttt{title\textbackslash ntext}. Confidence intervals are percentile bootstrap at $B=10^5$ throughout (the 1,000 questions are resampled with replacement $B=100{,}000$ times and the interval is the middle 95\% of the resulting values), marginal rows and paired families alike; the resampling unit is the evaluation question. Printed values use binary-float half-even rounding, so a displayed $x.xx5$ may round either way. \textbf{On Recall@5, only NV-Embed-v2 is statistically indistinguishable from the top entrant; every other embedder is significantly below it at 95\%, and that verdict survives multiplicity correction.} That verdict is Recall@5 only: the corrected family was computed at $k=5$ and no top-entrant-referenced Recall@10 family was run, so the @10 column carries marginal intervals and no multiplicity claim. It would not carry the same one: at $k=10$ the anchor sits \emph{above} the top entrant (78.12 vs 77.54), which is the sign reversal the abstract reports. Pairwise inference must not be read from the overlap of these marginal intervals: the paired comparisons, both correction families, and the simultaneous intervals are in \S\ref{sec:res-inference}. BAAI is the Beijing Academy of Artificial Intelligence.}
\label{tab:recall-panel}
\end{table*}

\begin{figure*}[t]
\centering
\includegraphics[width=0.92\textwidth]{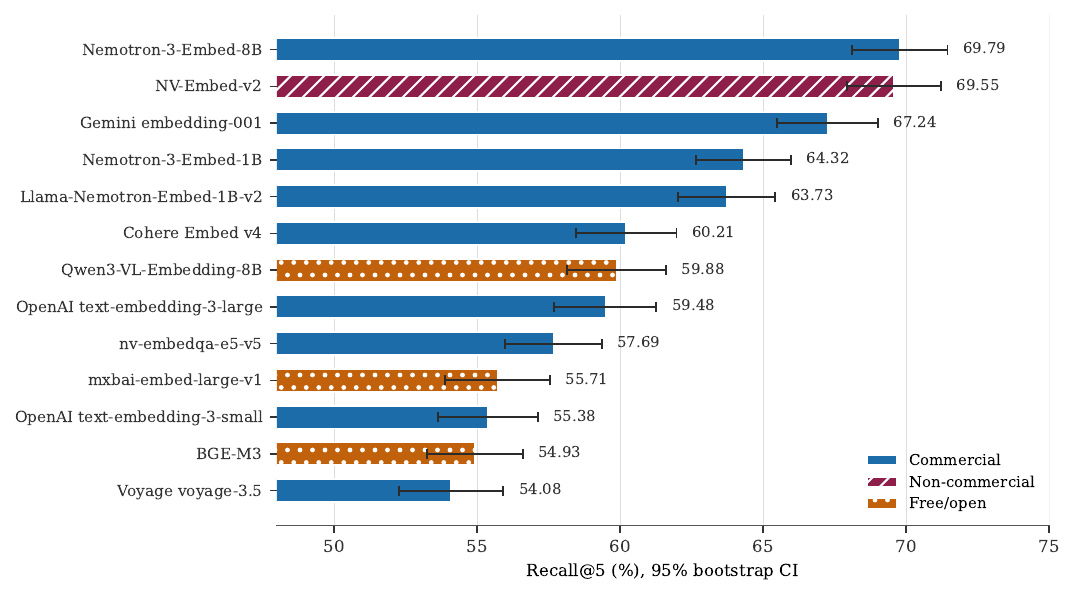}
\caption{Recall@5 across the panel with 95\% bootstrap confidence intervals, coded by license (hatching repeats the license coding so the figure survives greyscale printing). Nemotron-3-Embed-8B and NV-Embed-v2 overlap almost entirely, which is what the paired test reports ($p=0.69$). These are marginal intervals and show magnitude and precision only; pairwise separation must \emph{not} be read off their overlap. \S\ref{sec:res-inference} explains why, and the corrected paired comparisons in Table~\ref{tab:recall-panel} are the inferential statement.}
\label{fig:recall-panel}
\end{figure*}

All embedders, including NV-Embed-v2 itself, are measured on the identical harness with nothing layered on top (Table~\ref{tab:recall-panel}, Figure~\ref{fig:recall-panel}). We did not borrow NV-Embed-v2's number from the literature: query-formatting choice alone swings Qwen3-VL's score by 11.6 points, so a literature figure whose formatting we don't control is not safely comparable to our own best-variant measurements. Measured ourselves, NV-Embed-v2 scores 69.55 Recall@5, consistent with the 69.7 the protocol paper prints.

The headline: NVIDIA's commercial, self-hostable flagship now \emph{matches} the non-commercial research anchor. Nemotron-3-Embed-8B (69.79/77.54)\footnote{Throughout, a pair $x$/$y$ is (Recall@5, Recall@10) in percent.} sits 0.24 Recall@5 points above NV-Embed-v2's same-harness result (69.55/78.12) and 0.58 points below it on Recall@10; neither difference is significant, and because a failure to reject is not a demonstration of equivalence, we state what this design can and cannot resolve. At Recall@5 the paired difference is $+0.24$, 95\% CI $[-0.94, +1.43]$, $p=0.69$; at Recall@10 the anchor is nominally ahead by 0.58 points, the paired difference behind Table~\ref{tab:recall-panel}'s 77.54 against 78.12, so Nemotron-3-Embed-8B is at $-0.58$ (unrounded $-0.583$), 95\% CI $[-1.65, +0.48]$, $p=0.28$. Every anchor-referenced difference in this paper is signed entrant-minus-anchor; a negative number means the entrant trails. The top-entrant family (Table~\ref{tab:paired-intervals}) uses the opposite convention, reference-minus-entrant, so that a deficit behind the leader prints positive; Gemini appears there at $+2.55$ and here at $-2.31$, the same model against two different references. With $n=1{,}000$ the smallest difference this design could detect at 80\% power is roughly $1.7$ points at $k=5$, so intervals of this width cannot certify a gap smaller than that. A two-one-sided-test at a buyer-relevant margin of $\pm0.5$ points \emph{fails} at this sample size. How far it fails depends on an assumption that has to be stated. Taking the true difference as zero, certifying $\pm0.5$ at 80\% power needs $n\approx12{,}500$ (at a true difference of zero both one-sided tests must pass, so the power term is $z_{0.90}$, not $z_{0.80}$). Taking it as the observed $+0.24$, it needs $n\approx33{,}700$. The second is the one to quote against our own point estimate, and it is roughly 34x the questions we ran. The claim the data support is therefore that any residual tax is small relative to the 2.31 points measured before Nemotron-3-Embed-8B existed, not that it is provably zero (paired bootstrap over the same 1,000 questions: $+0.24$, 95\% CI [$-0.94$, $+1.43$], $p=0.69$; per question the two split 112 wins to 109 with 779 ties, \texttt{panel\_derived\_stats.json}). The two are statistically indistinguishable, and NV-Embed-v2 is the only entrant in the panel of which that is true.

The finding is that the gap has closed, not that there is a new winner. Before Nemotron-3-Embed-8B was released on 2026-07-16, the best commercially-deployable embedder in this panel was Google Gemini embedding-001 at 67.24, trailing the anchor by 2.31 Recall@5 points, a gap that \emph{is} significant (95\% CI [0.91, 3.71], $p=0.001$). A practitioner who needed a commercially-licensed retriever in June 2026 paid a real, measurable quality penalty for that requirement. As of July 2026 they no longer do. That is the commercial tax, and it closed within a single release cycle.

The durable finding, whichever embedder wins, is the paid-versus-free divide. The four API-only providers (OpenAI, Cohere, Google and Voyage AI) charge per token every time a corpus is re-indexed (\S\ref{sec:res-embed-cost}). A self-hosted open-weight model charges nothing per token, at any volume. (Voyage AI is the provider Anthropic's own documentation directs customers to, since \textit{``Anthropic does not offer its own embedding model''}~\cite{anthropic2026embeddings}.) Nemotron-3-Embed-8B is the only entrant that is simultaneously commercially licensed, free to self-host, and statistically indistinguishable from the anchor. The apache-2.0 and MIT entrants satisfy the license and self-hosting conditions but not the third. They dodge both the license restriction and the per-token toll, and pay for it in retrieval quality, sitting 9.7 to 14.6 points below the anchor.

Two secondary findings. First, Recall@10 lifts every embedder over Recall@5, by between $+7.62$ (Llama-Nemotron-Embed-1B-v2) and $+10.67$ (\texttt{text-embedding-3-large}), with a median of $+8.64$. Every one of the thirteen gains, and every one of the thirteen intervals around them, excludes zero, computed for all thirteen and not asserted as bare point estimates, which is the discipline \S\ref{sec:discussion} argues for. \emph{Paired} bootstrap means the two scores being compared are always recomputed on the \emph{same} resampled set of questions, so the comparison never benefits from one side happening to draw an easier sample. Narrowest as printed $[+6.75, +8.77]$ (tied at unrounded width with $[+6.62, +8.65]$), widest $[+9.48, +11.87]$. All thirteen, with the ten format-penalty intervals, are printed in Table~\ref{tab:paired-intervals}. The panel spans just over three points ($3.05$) despite covering eight makers and a 15-point spread in Recall@5. That consistency marks the pattern as a property of the task, not of any embedder, and motivates the k=5-vs-k=10 cost comparison of \S\ref{sec:res-f1}. Second, BGE-M3, one of the most widely-deployed open-weight embedders (35.6M HuggingFace downloads in the thirty days to 2026-08-15), lands second-to-last (dense mode; $54.93$ Recall@5 / $62.89$ Recall@10, the same pair reported for every entrant throughout). This is not a claim that BGE-M3 is weak in general; it is evidence that general-purpose leaderboard reputation does not transfer to a specific, stringently-filtered hard benchmark, which is exactly why this paper measures every embedder directly instead of citing reputation. (Format-sensitivity detail for both findings: Appendix~\ref{app:worked}.)

\subsection{Inference and Multiplicity}
\label{sec:res-inference}

A paired $p$-value here is the two-sided bootstrap $p$-value: twice the smaller of the two tail fractions of resampled differences, $2\min\{\Pr(\bar d^*\le 0), \Pr(\bar d^*\ge 0)\}$, with the usual $+1$ correction. Because it is a fraction of $B$ resamples it cannot go below $2/(B+1)$, which is why several $p$-values in this section print at that floor rather than at zero. The panel family and the headline pair use $B=10^5$; Table~\ref{tab:paired-intervals} states its own $B$.

What the intervals are intervals \emph{about} is worth one sentence, because the benchmark is a fixed set of 1,000 questions and not a probability sample of anything. Resampling those questions treats them as exchangeable representatives of a MuSiQue-like population of multi-hop questions, and the intervals accordingly quantify sensitivity to \emph{which questions were asked}. They do not quantify uncertainty across corpora, domains, or workloads: a second corpus could reorder the panel and nothing here would have predicted it. That is the estimand this paper's every $p$-value and confidence interval addresses.

Marginal overlap is not a paired test, and Figure~\ref{fig:recall-panel} is the place a reader is most likely to try it. Gemini's interval there overlaps both leaders by about a point, yet the paired tests separate it from the top entrant at $p=0.0003$ and from the anchor at $p=0.001$ ($B=10^5$; the latter is the pre-specified comparison the abstract quotes). The reason is that per-question scores across entrants are strongly correlated: $\rho=0.64$ on average across the panel, $0.75$ between the two leaders, both in \texttt{panel\_derived\_stats.json}. (The released artifact's \texttt{mean\_correlation}=0.44 is the correlation among the twelve \emph{difference} statistics, a different and smaller quantity.) So overlap of marginal intervals is a badly conservative and formally invalid test of a paired difference.

The claim that only NV-Embed-v2 is indistinguishable from the top entrant is a uniqueness claim over twelve pairwise comparisons against a common reference, so it is reported under Holm--Bonferroni correction at $\alpha=0.05$. At $B=10^5$, ten of the eleven $p$-values sit at the bootstrap's two-sided resolution floor ($p=2\times10^{-5}$; floored, not exact, and carrying no ordering among themselves\footnote{Ties at the floor were left in file order. No rejection depends on that choice: the smallest Holm threshold in the family, $\alpha/12=0.0042$, sits two orders of magnitude above the floor.}) and the eleventh, Gemini vs the top entrant, resolves to $p=0.0003$ against the eleven step-down thresholds $\alpha/12$ through $\alpha/2$ ($0.0042$ to $0.025$). Every rejection survives. The twelfth and last step tests NV-Embed-v2 at $\alpha/1=0.05$, which it fails at $p=0.69$, leaving it the sole entrant not separable from the top. Simultaneous Dunnett-type intervals give the same verdict: eleven of the twelve differences exclude zero jointly at 95\%, NV-Embed-v2 again the sole exception. The critical value is 2.78 against the marginal 1.96, and it is computed, not quoted. On each of $B=10^5$ resamples the questions are drawn once and jointly, so the twelve comparisons keep their real dependence. Each comparison is studentized against its own resampled standard error, and the critical value is the 95th percentile of the maximum absolute studentized difference across the twelve. The released \texttt{panel\_simultaneous\_ci.json} carries it (2.7821, against a Bonferroni 2.8653), with the twelve intervals and the seed. These simultaneous intervals were computed once, for the top-entrant family only (the artifact records \texttt{reference}: Nemotron-3-Embed-8B); the anchor-reference family is reported under Holm correction and without a parallel Dunnett-type band, so a reader checking the release should not expect to find a second set. An independent re-implementation written for this revision returns 2.7935 on the same vectors, seed and generator, a 0.4\% difference that comes from one implementation detail (which standard error each resample is scaled by) rather than from a different random draw; we report it rather than reconcile it away. The intervals printed in Table~\ref{tab:paired-intervals} are the release artifact's (\texttt{panel\_simultaneous\_ci.json}, 2.7821). The paired standard error behind this and the power calculations of \S\ref{sec:res-panel} is 0.60 Recall@5 points for the headline pair, from a per-question difference SD of 19.07 over $n=1{,}000$; every derived figure in this paragraph is regenerated by \texttt{scripts/panel\_derived\_stats.py} into \texttt{panel\_derived\_stats.json}.

Two comparison families are used in this paper, and both are corrected for multiple comparisons (Holm--Bonferroni, so that testing many pairs does not manufacture a significant one by chance). The first takes the top entrant as reference and is the one above. The second is its mirror, with the anchor as reference: it \emph{excludes} the pre-specified primary comparison and Holm-corrects the remaining eleven, whereupon all ten entrants other than Nemotron-3-Embed-8B reject (every $p$ at the $2\times10^{-5}$ floor against a smallest threshold of $\alpha/11=0.0045$) and Nemotron-3-Embed-8B is again the sole survivor ($p=0.69$). Both families are released as artifacts (\S\ref{sec:availability}). The primary Gemini-vs-anchor test stands alone at its pre-specified $p=0.001$, so ``only entrant indistinguishable from the anchor'' holds under correction in its own family, not merely by symmetry. That comparison, the June-2026 one, specified before Nemotron-3-Embed-8B entered our panel, is primary, and it is excluded from both corrected families by design, not by convenience: correcting a pre-specified primary test inside a family assembled afterwards would penalize it for company it never kept.

The Recall@10, lift, and format-penalty tests of \S\ref{sec:res-panel} and Appendix~\ref{app:reproduction} are secondary and reported uncorrected with that label. Correcting globally changes nothing. Under Holm over every paired test reported in this paper (a family of 47 tests: 12 top-entrant Recall@5 comparisons, plus 11 anchor-referenced ones not already among them (the primary Gemini-versus-anchor test is one of these), plus the Recall@10 headline pair, 13 Recall@10 lifts and 10 format penalties; $12+11+1+13+10=47$), all claimed rejections stand (the binding case, Gemini's format penalty at $p=0.0038$, ranks 44th and meets a step-down threshold of $\alpha/4=0.0125$) and \texttt{text-embedding-3-large}'s penalty still fails, exactly as claimed.

Note what correction cannot do. It disciplines the rejections, but the anchor's non-separability rests on the interval, minimum-detectable-effect and equivalence analysis above, not on a corrected $p$. Uncorrected, the family-wise error rate over twelve tests would be approximately 46\% under independence ($1-0.95^{12}$), and less under the strong positive correlation these twelve share through their common reference, which is why the correction is stated here.

\subsection{Literature Audit: License and Cost Disclosure}
\label{sec:res-audit}

License: 0 of the 3 systems relying on the anchor (HippoRAG-2, PropRAG, SAG) disclose NV-Embed-v2's non-commercial license anywhere: not in paper text, not in limitations, not in the linked code repository. The license is not hidden; it is stated on the model card. The finding is a failure to report a stated fact, not a failure to uncover a concealed one.

Cost: Table~\ref{tab:cost-disclosure}. Two of the five audited systems publish a dollar indexing figure (PropRAG $\sim$\$4, KET-RAG \$1.89, both on the identical 11,656-passage corpus); counts here and throughout are over the audited set, which \S\ref{sec:task-cd} describes and does not present as a random sample. HippoRAG-2 publishes token counts but no dollars: 9.2M input / 3.0M output to index the same corpus (its Table 12). The \$1.59 at gpt-4o-mini batch rates is \emph{our} conversion of that disclosure (Appendix~\ref{app:costnotes}), not a price its paper states. SAG and Microsoft's GraphRAG paper disclose no dollar figure at all, SAG despite claiming production deployment "at a scale of hundreds of millions of items."

\begin{table*}[t]
\centering
\footnotesize
\begin{tabular}{@{}L{2.6cm}L{2.6cm}L{9.5cm}@{}}
\toprule
System & Cost disclosed in own paper? & What the paper actually reports \\
\midrule
HippoRAG-2 & Tokens, not dollars & 9.2M input / 3.0M output tokens to index the corpus (its Table 12); \$1.59 at gpt-4o-mini batch rates is our conversion (Appendix~\ref{app:costnotes}), not a disclosed price \\
PropRAG & Yes & $\sim$\$4 to index the identical corpus (Appendix A.5) \\
SAG & No & Zero dollar figures despite qualitative cost discussion and claimed production scale; reports only a token-count-vs-recall curve not convertible to dollars \\
KET-RAG & Yes & \$1.89 to index the MuSiQue corpus with its own system (its Table 2, low-cost versions). That table --- captioned \textit{``Overall performance of RAG methods in low-cost versions''}, row \texttt{MS-Graph-RAG}, column \texttt{MuSiQue/USD} --- separately reports \$2.30 for Microsoft's Graph-RAG on that corpus --- the third-party figure used in Table~\ref{tab:indexing-cost}. Its introduction separately cites \$21 for knowledge-graph (KG-RAG) indexing of a 3.2MB HotpotQA sample, as motivation, not as KET-RAG's own cost \\
Microsoft GraphRAG (own paper) & No & Zero dollar figures; reports only wall-clock indexing time (281 minutes, one small dataset) \\
NVIDIA NeMo NIMs (platform) & License/infrastructure pricing only & \$1--8/hr AI Enterprise license tiers plus GPU compute --- structurally a different pricing model than per-corpus figures \\
\bottomrule
\end{tabular}
\caption{Cost-disclosure audit across the five audited systems. The sixth row, NVIDIA NeMo NIMs, is the hosting platform, not an audited system, and is listed for context only; it is excluded from the five-system counts in the abstract and \S\ref{sec:res-index-cost}.}
\label{tab:cost-disclosure}
\end{table*}

The spread in that row has a precise cause, and it is disclosure, not disagreement. KET-RAG prices GraphRAG twice on this exact corpus: \$2.30 to index it once under a ``low-cost'' configuration and \$24.94 under a ``high-performance'' one. That is an 11x range inside a single paper. Reconstructing the cost independently, from the token counts HippoRAG-2 reports for its own GraphRAG reproduction (115.5M input and 36.1M output, its Table 12, at the gpt-4o-mini batch and standard rates of \S\ref{sec:cost-model}), gives \$19.49--\$38.99.\footnote{KET-RAG's two figures are its Table 2, captioned \textit{``Overall performance of RAG methods in low-cost versions''} (input chunk size $\ell=1{,}200$), and its Table 3, \textit{``\ldots{}in high-performance versions''} ($\ell=150$); both are the \texttt{MS-Graph-RAG} row, MuSiQue \texttt{USD} column, each verified against the paper's arXiv \LaTeX{} source, not a rendering. The 11x is therefore a chunk-size choice: disclosed by KET-RAG, and absent from Microsoft's own paper, which is the point. The reconstruction converts HippoRAG-2's Appendix F counts (115.5M input / 36.1M output, Llama-3.3-70B) at gpt-4o-mini rates; it is directional only, since it crosses both tokenizers and models, and its own range is the batch-vs-standard tier spread. Detail in Appendix~\ref{app:costnotes}.} That range brackets KET-RAG's high-performance figure, so two independent estimates agree once the configuration is matched. What a reader cannot recover is \emph{which} configuration ``the cost of GraphRAG'' refers to, because Microsoft's own paper gives no dollar figure at all. An 11x build-cost spread that lives entirely inside an undisclosed configuration choice is this paper's finding in miniature.

\subsection{Indexing Cost at Realistic Scale}
\label{sec:res-index-cost}

\begin{figure*}[t]
\centering
\includegraphics[width=0.95\textwidth]{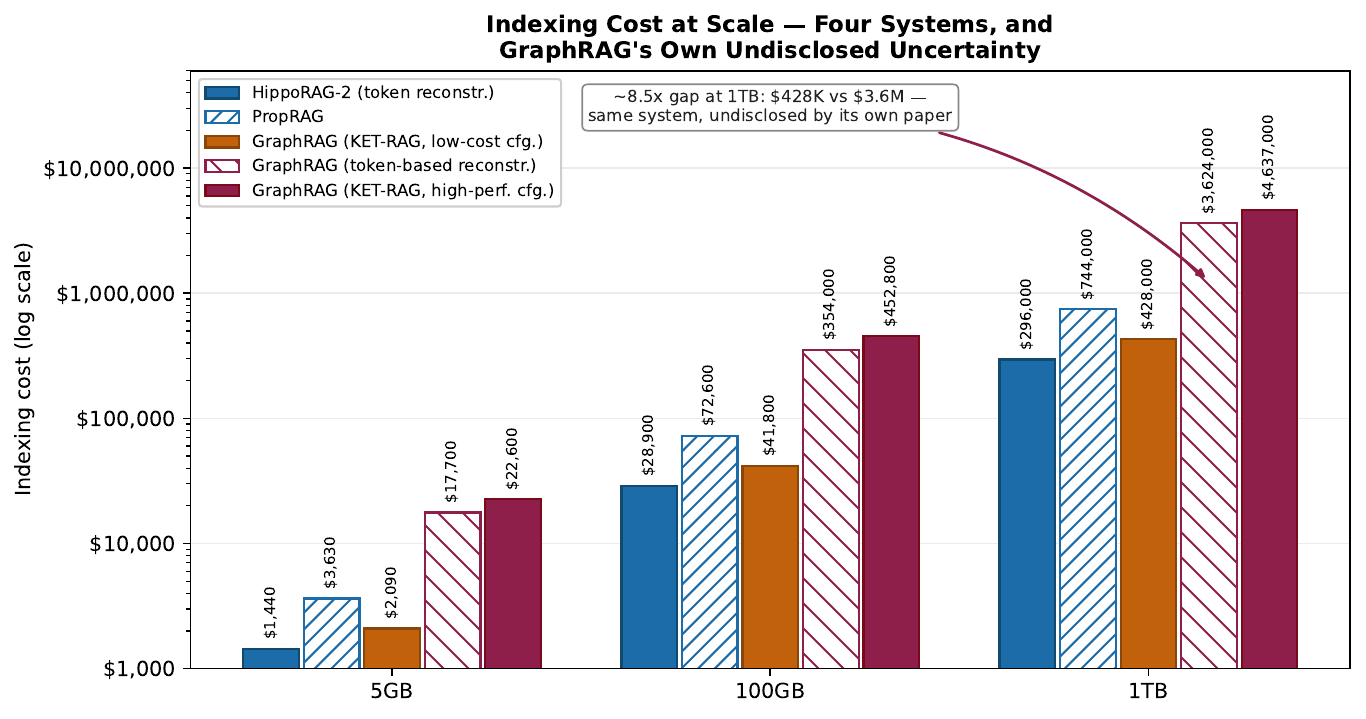}
\caption{Indexing cost at scale. GraphRAG's cost spans 11x between KET-RAG's low-cost and high-performance configurations (the ratio is scale-invariant under linear rates); the independent token-based reconstruction brackets the high-performance figure. A multi-million-dollar configuration choice at 1TB, about which GraphRAG's own paper discloses nothing.}
\label{fig:indexing-cost-scale}
\end{figure*}

Applying each system's own MuSiQue-derived \$/MB rate to three illustrative corpus sizes (Table~\ref{tab:indexing-cost}, Figure~\ref{fig:indexing-cost-scale}): the spread between GraphRAG's two configurations at 1TB (\$428K vs.\ \$4.6M) is the headline. It is a multi-million-dollar uncertainty at realistic scale, for a system whose own paper says nothing about dollars.

\begin{table*}[t]
\centering
\begin{tabular}{@{}lcccc@{}}
\toprule
System & Rate (\$/MB) & 5GB & 100GB & 1TB \\
\midrule
HippoRAG-2 (token-based reconstr.) & \$0.28 & $\sim$\$1,440 & $\sim$\$28,900 & $\sim$\$296,000 \\
PropRAG & \$0.71 & $\sim$\$3,630 & $\sim$\$72,600 & $\sim$\$744,000 \\
GraphRAG (KET-RAG, low-cost cfg.) & \$0.41 & $\sim$\$2,090 & $\sim$\$41,800 & $\sim$\$428,000 \\
GraphRAG (KET-RAG, high-perf.\ cfg.) & \$4.42 & $\sim$\$22,600 & $\sim$\$452,800 & $\sim$\$4,637,000 \\
KET-RAG (own system, low-cost) & \$0.34 & $\sim$\$1,720 & $\sim$\$34,300 & $\sim$\$351,000 \\
GraphRAG (token-based reconstr.) & \$3.46 & $\sim$\$17,700 & $\sim$\$354,000 & $\sim$\$3,624,000 \\
\bottomrule
\end{tabular}
\caption{Standardized indexing cost at three illustrative corpus sizes (linear extrapolation; see caveat in text). The \$/MB column is displayed rounded but every figure is computed from the unrounded rate (e.g.\ HippoRAG-2 is \$1.59/5.64MB $=$ \$0.2819/MB, from its disclosed 9.2M/3.0M indexing tokens at gpt-4o-mini batch rates: a reconstruction like the GraphRAG token row, not a published price). Sizes are binary in this section's cost tables: 1GB $=$ 1{,}024MB, 1TB $=$ 1{,}048{,}576MB (Appendix~\ref{app:tam} states its own, decimal, convention). The GraphRAG token-based row uses the \emph{low} end of the \$19.49--\$38.99 reconstruction of \S\ref{sec:res-audit}; its high end would put 1TB at \$7.2M. That range \emph{brackets} KET-RAG's high-performance configuration while sitting far above its low-cost one: GraphRAG's build cost is configuration-dominated, which is the finding.}
\label{tab:indexing-cost}
\end{table*}

One caveat, stated explicitly: these rates are extrapolated \emph{linearly}, which is almost certainly wrong for graph-construction systems; entity extraction and community detection plausibly scale super-linearly with graph density. If so, the 1TB figures are more likely \emph{under}estimates, which strengthens the argument. (KET-RAG's own \$21/3.2MB HotpotQA figure, and the provenance of its widely-repeated "\$33K for a 5GB legal case" extrapolation, are traced in Appendix~\ref{app:costnotes}.)

\subsection{Embedding Cost: One-Time, at Scale}
\label{sec:res-embed-cost}

Real published rates applied to the 11,656-passage MuSiQue corpus we measured at 5.64MB / $\sim$1.27M tokens, and extrapolated to a 10GB corpus (Table~\ref{tab:embedding-cost}): the four API providers charge \$46--\$346 per full re-index at 10GB at their standard tiers, recurring on \emph{every} re-index. Tier choice moves this range. OpenAI's batch API covers the embeddings endpoint at half price and Voyage documents a 33\% batch discount, so the cheapest 10GB re-index falls to $\approx$\$23 (text-embedding-3-small, batch) for a buyer who can wait hours. Cohere Embed v4 has no batch path on Bedrock (AWS's batch-inference support list carries no Cohere models~\cite{aws2026batch}; the only embedders on it are Amazon's own Titan and Nova), so its row is genuinely single-tier. The cheapest-provider ranking is therefore tier-sensitive, and Table~\ref{tab:embedding-cost} prints the tiers it prices; every self-hostable model carries no per-token charge, with the one-time hardware cost anchored at \$4,699 (\S\ref{sec:cost-model}). NV-Embed-v2 is also \$0. It is not legal to deploy commercially at any scale.

\begin{table*}[t]
\centering
\small
\begin{tabular}{@{}L{5.2cm}L{2cm}L{1.8cm}L{4.5cm}@{}}
\toprule
Embedder & Rate & MuSiQue corpus & @10GB \\
\midrule
OpenAI text-embedding-3-small & \$0.02/1M & \$0.03 & \$46 \\
Voyage voyage-3.5 & \$0.06/1M & \$0.08 & \$138 \\
Gemini embedding-001 (batch) & \$0.075/1M & \$0.10 & \$173 \\
Cohere Embed v4 (Bedrock) & \$0.12/1M & \$0.15 & \$277 \\
OpenAI text-embedding-3-large & \$0.13/1M & \$0.17 & \$300 \\
Gemini embedding-001 (standard) & \$0.15/1M & \$0.19 & \$346 \\
Nemotron-3-Embed-8B (self-host) & --- & \$0 & \$0, forever \\
Nemotron 1B tiers, \texttt{nv-embedqa-e5-v5} (NIM) & --- & platform-priced (Table~\ref{tab:cost-disclosure}) & platform-priced as measured; both 1B weight sets are public (Nemotron-3-Embed-1B: OpenMDW-1.1; Llama-Nemotron-Embed-1B-v2: NVIDIA Open Model License), so self-hosting either is \$0/token like the 8B. \texttt{nv-embedqa-e5-v5} has no public weights: its platform price is the only price \\
Qwen3-VL, mxbai, BGE-M3 & --- & \$0 & \$0, forever \\
NV-Embed-v2 & --- & \$0 & \$0 --- but not legal to deploy commercially \\
\bottomrule
\end{tabular}
\caption{One-time embedding cost on the 11,656-passage MuSiQue corpus (5.64MB of raw text, $\sim$1.27M tokens, $\sim$1.6\% of them evaluation queries), and extrapolated to a 10GB corpus. Token density here is empirical to this corpus ($\approx$4.66 characters/token, queries included, with 5.64MB read as binary per the Table~\ref{tab:indexing-cost} convention). Appendix~\ref{app:tam} instead applies a 4-characters/token approximation to legal text, so its 1TB embedding figures run about 10\% above what size-scaling this table would give; the two are not meant to be scaled into one another.}
\label{tab:embedding-cost}
\end{table*}

\subsection{Answering Cost: Recurring, per Query}
\label{sec:res-f1}

We ran gpt-4o-mini as a fixed LLM over the real top-$k$ retrieved passages, for all 1,000 questions, at both k=5 and k=10, measuring real API cost (Table~\ref{tab:answer-cost}).

\begin{table*}[t]
\centering
\footnotesize
\begin{tabular}{@{}lcc@{}}
\toprule
Embedder & Cost @k=5 (batch/std) & Cost @k=10 (batch/std) \\
\midrule
Nemotron-3-Embed-8B & \$0.058 / \$0.116 & \$0.112 / \$0.225 \\
Gemini embedding-001 & \$0.057 / \$0.113 & \$0.108 / \$0.215 \\
Nemotron-3-Embed-1B & \$0.055 / \$0.110 & \$0.105 / \$0.209 \\
Llama-Nemotron-Embed-1B-v2 & \$0.059 / \$0.118 & \$0.114 / \$0.229 \\
OpenAI text-embedding-3-large & \$0.058 / \$0.116 & \$0.108 / \$0.217 \\
Cohere Embed v4 & \$0.056 / \$0.113 & \$0.107 / \$0.215 \\
Voyage voyage-3.5 & \$0.055 / \$0.110 & \$0.103 / \$0.206 \\
Qwen3-VL & \$0.053 / \$0.105 & \$0.099 / \$0.198 \\
\texttt{nv-embedqa-e5-v5} & \$0.055 / \$0.109 & \$0.103 / \$0.206 \\
mxbai & \$0.057 / \$0.115 & \$0.108 / \$0.216 \\
OpenAI text-embedding-3-small & \$0.058 / \$0.117 & \$0.110 / \$0.219 \\
BGE-M3 & \$0.060 / \$0.120 & \$0.112 / \$0.223 \\
\bottomrule
\end{tabular}
\caption{Answering cost at k=5 and k=10, all 1,000 queries, gpt-4o-mini as the LLM, at batch and standard API rates. Twelve rows, not thirteen: NV-Embed-v2 is absent because answering cost depends on the retrieved passages' token length, and the anchor's answering run was not repeated after the index correction. Its cost would fall inside the range shown: the twelve span \$0.053--\$0.060 at k=5, a spread of \$0.007 per 1,000 queries, and the spread, not the cost, is the small number here. We do not report a figure we did not measure.}
\label{tab:answer-cost}
\end{table*}

Two observations. First, applying the published gpt-4o-mini rates of \S\ref{sec:cost-model} to the k=5 context length gives \$0.055--\$0.12 per 1,000 queries across the two rate tiers; measured, the batch range is \$0.053--\$0.060 and the standard range \$0.105--\$0.120, so the rate card predicts the meter on both tiers to within 4\%. Second, the k=5$\rightarrow$k=10 cost ratio is tightly clustered across the panel: mean 1.89x, range 1.86--1.93x. The ratio is tier-invariant, because answering cost is driven by how much retrieved text the LLM is handed, not by which embedder retrieved it. (The standard tier prices every cell at exactly twice the batch tier before rounding, so the doubling cancels; the slightly different range a reader would recompute from the printed standard column, up to 1.94x, is an artifact of rounding those cells to the nearest \$0.001, not a real tier effect.) Doubling the passages from five to ten does not quite double the cost (the ratio sits below 2x) because part of each request does not scale with $k$: the question, the prompt instructions, and the generated answer are the same size whether five passages or ten are attached. Passage length is fixed by the benchmark here (MuSiQue's Wikipedia passages average 79.8 words), so this ratio is a property of that corpus; a deployment that chunks its own documents differently would see a different constant.

Scaled to an illustrative production volume of 10,000 queries/day for a year, the measured k=10 answering cost extrapolates to \$361--\$836/year (Qwen3-VL at batch rates to Llama-Nemotron-Embed-1B-v2 at standard rates). How that compares to the one-time costs is not the intuitive ordering. A single 10GB re-index costs \$46--\$346 on an API embedder (\S\ref{sec:res-embed-cost}) and \$0 on a self-hosted one, so at this query volume the \emph{recurring} answering line overtakes the \emph{one-time} embedding line inside the first year, and keeps going. The crossover a budget-holder should carry: one-time API re-embedding of the corpus equals one year of answering at roughly 10--180GB of corpus, both bounds using the standard embedding tiers of Table~\ref{tab:embedding-cost}. The two ends are not symmetric and it is worth saying which is which: the 10GB end is the cheapest answering line (\$361/yr, Qwen3-VL at batch) against the dearest standard embedding rate (\$34.6/GB), and the 180GB end is the dearest answering line (\$836/yr, at standard) against the cheapest embedding rate (\$4.6/GB). Holding the answering side to standard tiers on both ends would move the lower bound to $\approx$21GB. Separately, on the embedding side, the upper bound roughly doubles to $\approx$360GB if the cheapest batch tier is used; below that range the recurring line dominates, far above it the one-time line does. Neither is the number that decides a budget, but the gap is smaller than a casual reading suggests, and it must be quoted at a matched corpus size to mean anything. At 1TB, the scale at which the graph-construction figures run to six and seven figures, API embedding is \$5.2K--\$39.3K (Appendix~\ref{app:tam}, Table~\ref{tab:tam-dollars}) against indexing of \$296K--\$4.64M. That is a factor of roughly 7.5 to 900, not the three-to-four orders of magnitude a 10GB-versus-1TB comparison would imply. A \$35,000 re-index is not a rounding error to a CFO; it is a budget line that happens to sit under a much larger one. The practical reading is that embedding choice is a licensing and sovereignty question rather than a cost question, answering cost is a volume question, and indexing architecture is where the money actually is. Disclosure runs inversely to size: of the three cost lines, the largest is the one the literature reports least.

\section{Discussion}
\label{sec:discussion}

\subsection{The Commercial Tax Was Real, Has Just Closed, and Was Never Academic}
\label{sec:disc-tax}

The tax was measurable, and it was recent. In June 2026 the best commercially-licensed retriever in this panel was Gemini embedding-001, at 67.24 Recall@5 against the field's non-commercial anchor at 69.55. That is 2.31 points, with a 95\% confidence interval of [0.91, 3.71] that excludes zero. It is a claim about the thirteen entrants measured here, not a survey of everything on the market; the pivot is Nemotron-3-Embed-8B's release on 2026-07-16~\cite{nvidia2026nemotron8b}. It is not a rounding error dressed up as a finding. It is a real quality penalty attached to a licensing requirement, and every commercially-deployable option in our panel paid it. Whether two points of Recall@5 matter is a question about the decision downstream of the retrieval, not about the metric. Where the passages a system surfaces are the evidence for committing capital, pricing a risk, or a national-security judgement, the gap is not a leaderboard position. It is the average shortfall in supporting evidence per query. And because a query that loses one of four gold passages contributes only a quarter of its weight to that average, the share of queries affected is larger than the mean suggests. On the released per-question vectors the commercial reference surfaces less of the gold than the anchor on 184 of the 1,000 questions and more on 113, with 703 ties. How many downstream decisions those misses would have changed is not something this design measures. What it can say is that the rate is paid on every query for as long as the system runs. One month later, Nemotron-3-Embed-8B is statistically indistinguishable from the anchor and the penalty is no longer measurable at this sample size. That is a weaker statement than "gone." The weaker one is what the data support (\S\ref{sec:res-panel}).

What closed, and what did not. The tax closed \emph{at the top}. Exactly one commercially-licensed embedder reaches the anchor, and it was released eight days before this paper's initial draft was finalized (2026-07-16 against 2026-07-24). The other eleven entrants remain 2.55 to 15.71 points behind the top entrant, all significantly so. A practitioner who picks a commercial embedder without checking which one still pays the tax in full. "The commercial tax has closed" is a statement about the frontier. It is not a statement about the category.

At the raw-embedder level, the free open-weight alternatives (Qwen3-VL, mxbai, BGE-M3) trail NV-Embed-v2 by 9.7--14.6 Recall@5 points. That is far more than the 1--3 points a system-level comparison suggests once reranking or graph structure compensates on top. A leaderboard that only ever showed system-level results with NV-Embed-v2 baked in quietly asserted that the embedder choice barely mattered. At the floor it is one of the largest single levers in the pipeline, and as of mid-2026 no longer one that forces a buyer into a non-commercial corner. BGE-M3's result sharpens the same point from the other side. General-purpose reputation is not a proxy for performance on the specific hard benchmark that matters to a specific buyer.

A methodological point falls out of our own error. While re-running the panel to attach confidence intervals, we found that the anchor's index had been built from body text alone while every other entrant had title and text together, a one-line defect that understated the anchor by 2.5 points in the direction that favored our thesis. Corrected, the anchor landed at 69.55, within 0.15 of the 69.7 its protocol paper prints, and our headline verb changed from ``exceeds'' to ``matches.'' A field that reports point estimates without intervals, and releases neither harness nor per-item outputs, has no mechanism by which an error of this shape would ever surface. The missing numbers are not merely inconvenient for buyers; they are what makes the published ones checkable.

Nor is the gap academic. Organizations build on retrieval instead of fine-tuning precisely because it is cheaper to keep current and easier to audit (\S\ref{sec:intro}). If so, the retrieval layer's own licensing and cost profile is the thing the entire adoption argument rests on. A benchmark that reports "82\% Recall@5" without saying whether it is legal to deploy, or what it costs to reproduce, answers a question practitioners are not asking, with a number they cannot use.

The timing is notable. On the day the initial draft of this paper was finalized, 2026-07-24, twenty-five organizations, NVIDIA and the Linux Foundation among them, published \textit{Open Weights and American AI Leadership}, arguing that open-weight models let organizations control cost, avoid lock-in, and retain data sovereignty \cite{openweights2026}. Our paid-versus-free divide is that letter's cost argument with dollar figures attached. Our results also sharpen a distinction the letter does not draw: \emph{open-weight is not the same as commercially deployable}. NV-Embed-v2 is exactly as downloadable as any model the letter celebrates, and legally unusable in a commercial product. (The letter is an advocacy document. We cite it as evidence of salience, not support.)

\subsection{Why Expensive-But-Governed Might Be Correct}
\label{sec:disc-governed}

The cost audit and the standardized cost table together make an underappreciated case. Graph-augmented and SQL-structured retrieval are not free, are frequently undisclosed, and at realistic scale the undisclosed number can plausibly reach millions of dollars. We are precise about what this does and does not claim: we have no Recall or cost figures for any specific commercial structured-retrieval product. The argument is conditional. If structured, governed retrieval carries a cost premium of the shape we measured for GraphRAG-category systems, then the premium buys auditability and governance, and whether it is worth paying is a requirements question for the buyer, not one this paper's data can answer. Whether "realistic scale" is itself real is established in Appendix~\ref{app:tam}, where we measure a single legal-diligence data room from a public benchmark at 225MB of extracted text. That is forty times this paper's entire MuSiQue corpus, for one deal.

\subsection{What a Buyer Should Ask}
\label{sec:discussion-buyer}

The findings above reduce to four questions a purchaser can put to any retrieval vendor, none of
which requires access to the vendor's source code and each of which this paper shows is currently
unanswered in the published literature.

\begin{enumerate}
\item \textbf{\emph{What is the license on the retrieval backbone, and does it permit commercial
deployment?}} Not on the system; on the embedding model inside it. None of the three audited systems whose numbers rely on it disclose that those numbers depend on a \texttt{cc-by-nc-4.0} model
(\S\ref{sec:res-license}). A system whose quality figure cannot legally be reproduced in
production is quoting a number the buyer cannot buy. This paper is not legal advice: where a model card or training-data lineage indicates non-commercial terms, deployability should be confirmed with counsel and the vendor before a benchmark result is treated as production-transferable.

\item \textbf{\emph{What did indexing cost on a corpus the size of my organization's, and is that
one-time or recurring?}} The three cost lines have different shapes (\S\ref{sec:res-index-cost}): graph
construction is the largest and is paid per rebuild; embedding is small, linear, and \$0 per token
if the model is self-hosted; answering scales with queries, not with corpus size. A single blended
figure hides which of the three will dominate a given deployment.

\item \textbf{\emph{Is the embedder self-hostable, or is there a per-token toll on every re-index?}} This is
the durable divide (\S\ref{sec:res-embed-cost}) and it does not close with the next model release.
Five of the thirteen embedders measured here are third-party API-only, from four providers, and a sixth, NVIDIA's \texttt{nv-embedqa-e5-v5}, is served only as a licensed NIM with no free self-host path; re-indexing a corpus after a schema
change, a chunking change, or a model upgrade re-pays that toll in full, every time.

\item \textbf{\emph{Was the quality figure measured on the same corpus format, the same $k$, and the
same protocol as the comparison?}} Corpus format alone moves Recall@5 by up to 4.73 points on this benchmark and query formatting moves one entrant by 11.6 (Appendix~\ref{app:worked},
Appendix~\ref{app:reproduction}). Numbers carried across harnesses are not comparable, and the
paper reporting them rarely says which harness produced them.
\end{enumerate}

None of these is a research question. They are diligence questions, and the reason they are hard to
answer today is that the reporting norm in this literature does not require the answers.

\subsection{What This Paper Leaves for Future Work}
\label{sec:disc-future}

Two questions are out of scope, because each requires introducing a specific retrieval system (ours), which this paper does not do. First, how much of the raw embedder gap a well-built system recovers through query-time machinery layered on the embedder, now that a commercial embedder wins the floor test. Second, the LLM-tier tradeoff: what accuracy a 2B, 70B, or frontier LLM buys per dollar. \S\ref{sec:res-f1} showed that a \emph{fixed} LLM's cost is small and the k=5-versus-k=10 choice cheap. It does not show what a \emph{better} LLM's cost buys. That is the subject of a planned follow-on paper.

\section{Limitations}
\label{sec:limitations}

\begin{itemize}[leftmargin=1.8em]
\item Three of NVIDIA's 12+ embedding NIM endpoints appear in the panel (Nemotron-3-Embed-1B, Llama-Nemotron-Embed-1B-v2, \texttt{nv-embedqa-e5-v5}), chosen for scale and architecture relevance. Three further identifiers were not served under our credentials at measurement time: \texttt{nv-embedqa-mistral-7b-v2}, no longer listed in the NIM support matrix, plus \texttt{nemotron-3-embed-8b} and \texttt{nemotron-3-embed-1b-bf16}, whose identifiers are specific to other distribution channels (detail in Appendix~\ref{app:costnotes}). Six endpoints attempted, three served.
\item BGE-M3 tested in dense-vector mode only, for comparability with the all-dense panel; its hybrid dense+sparse+multi-vector mode was not tested and could plausibly close some of its gap, though closing 12--15 points on this benchmark is not something we would assume without measuring.
\item The GraphRAG cost spread is explained (a chunk-size choice, \S\ref{sec:res-audit}) but not resolved: Microsoft's own paper names neither configuration, so which figure "the cost of GraphRAG" should mean is a choice the reader has to make. We report both.
\item All rates are as published 2026-07-21 and subject to provider change; the \emph{relative} structure (self-hosted free forever, API recurring, indexing dominant) should hold regardless.
\item One benchmark, one corpus. Everything here is measured on MuSiQue's 11,656-passage corpus under the HippoRAG-2 protocol, so every number should be read with that scope attached. We ran a small pilot on 2WikiMultihopQA with three models (the anchor, the winning entrant and the June-2026 commercial reference), on the same harness and the same released question-file convention, with per-question vectors released as \texttt{probe\_nvembed\_2wikimultihopqa.json}, \texttt{probe\_nemotron\_2wikimultihopqa.json} and \texttt{api\_panel\_sweep\_2wikimultihopqa\_perq.json}. It separates the two claims this paper makes. The \emph{match} replicates: Nemotron-3-Embed-8B versus the anchor is $-0.30$ (95\% CI $[-1.23, +0.62]$, $p=0.54$) at Recall@5 there, against $+0.24$ ($p=0.69$) here, the sign reversing between corpora as one expects of two systems that are level. The \emph{magnitude of the tax} does not: the commercial reference point that trails the anchor by 2.31 points on MuSiQue trails it by $0.12$ (95\% CI $[-1.07, +0.82]$, $p=0.82$) on 2Wiki, which is no measurable gap at all. That weakens the magnitude claim, and we report it because a paper about undisclosed measurements should not hold one back. It also narrows the claim to what the evidence supports: the tax was real and is now closed \emph{on this corpus}, and how large it was in the first place is a property of the corpus, not a constant of the field.
\item Our cost model prices machine time (LLM inference and embedding) and not human time. Setting up a structured or graph-augmented retrieval system plausibly involves schema, ontology and prompt engineering by people, none of which we attempted to price and none of which our disclosure audit looked for: that audit searched each paper for currency markers tied to indexing, so it can report the absence of dollar figures but not the absence of any particular cost category. Whether human setup cost is material at deployment scale is an open question this paper does not answer.
\item Benchmark retrieval is exact and production retrieval is not (\S\ref{sec:cost-model}). Every Recall@$k$ here is the index-free ceiling for its embedder; a deployed ANN index gives some of it back, and how much is a property of that index, not of the model, so we do not estimate it. The comparison between embedders is unaffected, since none of the thirteen is indexed. But no number in this paper should be read as a production recall figure.
\item Our pre-specification of the primary comparison is asserted in this text and was not registered externally before the runs. Our own version history dates the commercial-tax question, best commercial entrant against the anchor, to 2026-07-18, and Nemotron-3-Embed-8B's entry into the panel to 2026-07-22; both are after its 2026-07-16 release, so "before it entered our panel" is what the record supports, and "before it existed" is not. A reader who wants that guarantee cannot get it from us; what they can get instead is every per-question vector, from which both the pre-specified family and the alternative are recomputable (\S\ref{sec:res-inference}), so the choice can be audited even though the timing cannot.
\item We measure our own fixed LLM's answering cost, not each audited system's query-time retrieval-LLM cost (entity linking, beam search, SQL generation). None of the four systems disclose that figure either, and we could not reconstruct it responsibly. This is a real gap in the cost accounting, and we state it.
\item Corpus format is directly verified for all thirteen entrants, ten by re-measurement under both formats (Appendix~\ref{app:reproduction}) and three (Qwen3-VL-Embedding-8B, mxbai-embed-large-v1 and BGE-M3) whose corpus embeddings were built ahead of this work and reused from cache. For those three we re-embedded the same 120-passage sample of the corpus in both forms with the same model and settings that embedded their queries (for Qwen3-VL the hosted open-weight instance at 1,024-dimensional truncation with no instruction; for the other two the local models as released) and compared against the cached rows: \texttt{title\textbackslash ntext} matches at cosine 1.0000 on every one of the 120 for all three, and text-only does not (medians 0.963, 0.989 and 0.987), so their format is confirmed, not inferred. What we did not do is rebuild those three corpora in text-only form for the format-penalty appendix, which is why their text-only cells in Table~\ref{tab:reproduction} are absent, not measured; no format penalty is claimed for them. Given that the defect we found affected exactly one model and was invisible in aggregate, we state exactly which cells are measured and which are not, instead of implying uniform coverage.
\item Eight of the thirteen entrants were measured through hosted endpoints exposed to silent server-side change: five API-only models (of which Gemini is the one pinned to a stable, non-preview release) and the three NIM-served NVIDIA tiers, so their weights may change under a stable model name. \texttt{text-embedding-3-large} moved $-0.54$ Recall@5 between our two measurement rounds, the largest such drift we observed. Hosted figures in this paper are therefore point-in-time in a way the self-hosted ones are not.
\end{itemize}

\section{Conclusion}
\label{sec:conclusion}

RAG is the dominant architecture industry uses to connect large language models to proprietary data, adopted specifically because it is supposed to be cheaper and more auditable than the alternative. We show that the multi-hop benchmarks the field uses to rank itself leave out the two things a real buyer needs to know. On licensing: none of the three systems relying on the anchor discloses that their retrieval backbone is non-commercial, a restriction we trace to a specific training dataset. On cost: of five systems, two publish no dollar indexing figure at all, one publishes token counts but no dollars, and for one of the two silent systems the only reconstructable figures, both third-party, span 11x with the configuration choice undisclosed, a multi-million-dollar uncertainty at realistic scale.

Measured at the embedder level on one harness, the cleanest test that requires no system of our own, a commercial, self-hostable embedder is now statistically indistinguishable from the non-commercial research anchor on retrieval. On this corpus the quality penalty for requiring a commercial license was a significant 2.31 Recall@5 points as recently as June 2026 and is now not measurable. We are deliberate about the verb: Nemotron-3-Embed-8B \emph{matches} the anchor, it does not beat it, and it is the only one of twelve commercially-deployable entrants that does. The durable finding is not "NVIDIA wins" but the paid-vs-free divide: API providers carry a permanent per-token toll, while every self-hostable open-weight model costs nothing per token in provider charges once embedded. Both embedding and answering costs are real money, and both sit far below the six-to-seven-figure graph-construction costs this literature never discloses: embedding by roughly 7.5x to 900x at matched 1TB scale, a year of answering at 10,000 queries/day by 350x or more. The two carry different multipliers on purpose (\S\ref{sec:res-f1}): the second is an annual recurring sum set against a one-time build. That is the context in which a buyer can weigh "expensive but governed" retrieval against its alternatives with the costs actually on the table.

\paragraph{What this measures, and what comes next.} One corpus is what we measured, and it is not what we claim. The generalization a buyer actually wants, \emph{will a self-hostable specialist hold up on my documents, in my domain?}, is not answerable from any public benchmark. The reason cuts against the frontier models, not for them. Public multi-hop corpora are built from Wikipedia, which sits inside the training data of every frontier system that reports on them. A model with hundreds of billions of parameters brings a great deal of memorized Wikipedia to a Wikipedia benchmark, and that advantage is not available on a corpus it has never seen. Our expectation for the follow-on work is therefore asymmetric and falsifiable: on public corpora a compact self-hostable embedder should sometimes match the frontier and sometimes trail it, but should rarely exceed it; on private corpora, where memorization contributes nothing and the specialist can be adapted to the domain, the ordering should be more favourable. Our 2Wiki pilot (\S\ref{sec:limitations}) is consistent with the first half of that and shows the gap's magnitude is a property of the corpus, not a constant. The next paper in this series tests the claim properly, across 2WikiMultihopQA and FRAMES and on a private corpus, with the answering LLM varied as well as the embedder, because the interesting hypothesis is not that one model wins but that there is no one-size-fits-all model, and that the size which wins depends on whose documents are being searched.

Testing the private half requires a corpus that no model has trained on, and no public benchmark can be that. We are building one: multimodal, with audio and imagery among the modalities precisely because they sit outside the text crawls these models were trained on. We state two things about it in advance. It will be released, because a benchmark authored by a party with a stake in the outcome is worth very little unless others can run their own models on it and check the result. That is the same standard we applied to the five papers audited here. And the private-corpus prediction above is a hypothesis registered before the measurement exists, not a finding: if a frontier model matches a domain-adapted specialist on a corpus it has never seen, that is the result we will report.

\paragraph{Competing interests, and what follows.} This paper measures the embedding backbone alone, and the authors have a stake in the result: we are building retrieval systems we intend to commercialise, and a follow-on paper will report an end-to-end system with an LLM-quality bake-off and energy accounting. Readers are entitled to ask whether that interest shaped the finding, so we say where it could not have, and where it might.

It could not have shaped the panel results, for a mechanical reason. Every entrant ran on one harness, with the corpus format, query mode, decoding parameters and scoring code fixed in advance and applied identically; the per-question vectors behind every number are released, so the ranking is recomputable by anyone without re-embedding a corpus. The headline finding is also not a win for the model we would prefer to promote. Nemotron-3-Embed-8B \emph{matches} the non-commercial anchor (+0.24, 95\% CI $[-0.94, +1.43]$, $p=0.69$) and we decline to call that a victory, which is the opposite of what a vendor writing marketing would do.

Where the interest is real is in \emph{selection}: we chose to run this audit on the licensing question rather than another, and a commercially licensed winner is convenient for us. We cannot neutralise that with method, only disclose it. The same applies, more sharply, to the corpus announced above: we are building the benchmark on which we predict our own class of system will do comparatively well, and no amount of care makes that a neutral act. The only real protection is the one we have committed to: releasing it, so that anyone who suspects the corpus was shaped to the conclusion can run a frontier model on it and say so. The reader's protection is the released harness. Run it against an embedder we omitted and the panel either holds or it does not.

\section*{Code and Data Availability}
\label{sec:availability}

A paper arguing that this literature under-discloses is obliged to publish its own harness. All measurement code and every result artifact behind Table~\ref{tab:recall-panel}, Table~\ref{tab:answer-cost} and Table~\ref{tab:reproduction} are released at \ARTIFACTREPO{} and archived, immutably and with a persistent DataCite DOI, at Zenodo: \ARTIFACTDOI{} (deposited \ARTIFACTDATE{}). The DOI, not the repository URL, is the citable record: a repository can be renamed, force-pushed or deleted, and a paper that under-discloses its own artifacts would be an odd vehicle for the argument made here.

The release includes the retrieval harness for all thirteen embedders, the answering-LLM evaluation script with its exact prompts and decoding parameters, the standardized cost model, and the bootstrap code. Both comparison families are released as computed artifacts, not as numbers printed only in this paper: \texttt{panel\_multiplicity.json} (reference = the top entrant) and \texttt{panel\_anchor\_family.json} (reference = the anchor, produced by \texttt{scripts/panel\_anchor\_family.py}), the latter carrying the primary Gemini-vs-anchor test quoted in the abstract and the Recall@10 leader pair, at the same $B$ and seed as the former. The simultaneous intervals and their max-$t$ critical value ship the same way, as \texttt{panel\_simultaneous\_ci.json} from \texttt{panel\_simultaneous\_ci.py}, so that no inferential number in this paper is one a reader has to take on trust. Independent readers of the previous drafts asked, in turn, where each of these could be checked; all of them can now be recomputed from the released vectors in one command each, which is the answer this paper's own argument requires. Critically it also includes the \emph{per-question} recall vectors for every entrant, one value per evaluation question in corpus order, so that every confidence interval and paired test in this paper can be recomputed, and so that any reader can run a comparison we did not think to run without re-embedding a corpus. The large embedding matrices exceed Zenodo's per-record ceiling and are mirrored as a HuggingFace dataset at \ARTIFACTHF{}, pinned to a revision hash, not a bare dataset name, so that the mirror is as fixed as the archive.

We publish the defective indexing script alongside the corrected one, unmodified. The MuSiQue corpus itself is not redistributed; the release documents how to fetch it from the upstream HippoRAG-2 project.

\section*{Disclosure of AI Assistance}

The authors used Gemini 3.6 Flash and Claude Opus 5 for drafting, editing, code review, and figure refinement; all scientific content, methodology, analysis, and conclusions were developed and verified by the authors, who take full responsibility for the paper.

\appendix

\section{Panel Composition}
\label{app:panel}

\begin{table*}[t]
\centering
\footnotesize
\begin{tabular}{@{}L{4.0cm}L{2.9cm}L{8.3cm}@{}}
\toprule
Embedder & Access & Notes \\
\midrule
NV-Embed-v2~\cite{nvidia2024nvembed} & Self-hosted & Non-commercial research anchor, measured on this harness, not taken from the literature (\S\ref{sec:res-panel}) \\
Nemotron-3-Embed-8B-BF16~\cite{nvidia2026nemotron8b} & Self-hosted & OpenMDW-1.1, "ready for commercial use." Released 2026-07-16; not served by NVIDIA's hosted NIM endpoint \\
Nemotron-3-Embed-1B-BF16~\cite{nvidia2026nemotron1b} & NIM API (served id \texttt{nemotron-3-embed-1b}) & Cheaper Nemotron tier \\
Llama-Nemotron-Embed-1B-v2 & NIM API & Cheaper Nemotron tier \\
nv-embedqa-e5-v5~\cite{nvidia2026e5} & NIM API & 335M params, fine-tuned E5-Large-Unsupervised, trained on NVIDIA's own commercially-viable QA data \\
text-embedding-3-small / -large~\cite{openai2026} & Direct API & --- \\
Cohere Embed v4~\cite{cohere2026} & Amazon Bedrock & Native foundation model; Cohere is an independent Canadian company, accessed via Bedrock but not owned by Amazon \\
Gemini embedding-001~\cite{google2026} & Direct API & Pinned to the stable (non-\texttt{-preview}) release for reproducibility \\
voyage-3.5~\cite{voyage2026} & Direct API & Voyage AI is a subsidiary of MongoDB, Inc. \\
Qwen3-VL-Embedding-8B & Self-hosted & Alibaba Cloud (Tongyi Lab), \texttt{apache-2.0}; free, open-weight baseline \\
mxbai-embed-large-v1 & Self-hosted & Mixedbread AI GmbH (Germany), \texttt{apache-2.0}; free, open-weight baseline \\
BGE-M3 & Self-hosted & BAAI, \texttt{mit}. Multilingual/multi-granularity (100+ languages, 8192-token context, dense+sparse+multi-vector); dense mode only tested here, for comparability \\
\bottomrule
\end{tabular}
\caption{Panel composition: access channel and per-model notes. Maker and license for each entrant are in Table~\ref{tab:recall-panel}.}
\label{tab:panel}
\end{table*}

We did not include an Anthropic/Claude embedder because none exists: Anthropic does not ship a text-embedding model, and its documentation recommends third-party providers, principally Voyage AI, already in this panel as voyage-3.5.

\section{Worked Metric Example and Format Sensitivity}
\label{app:worked}

\begin{figure*}[t]
\centering
\includegraphics[width=0.95\textwidth]{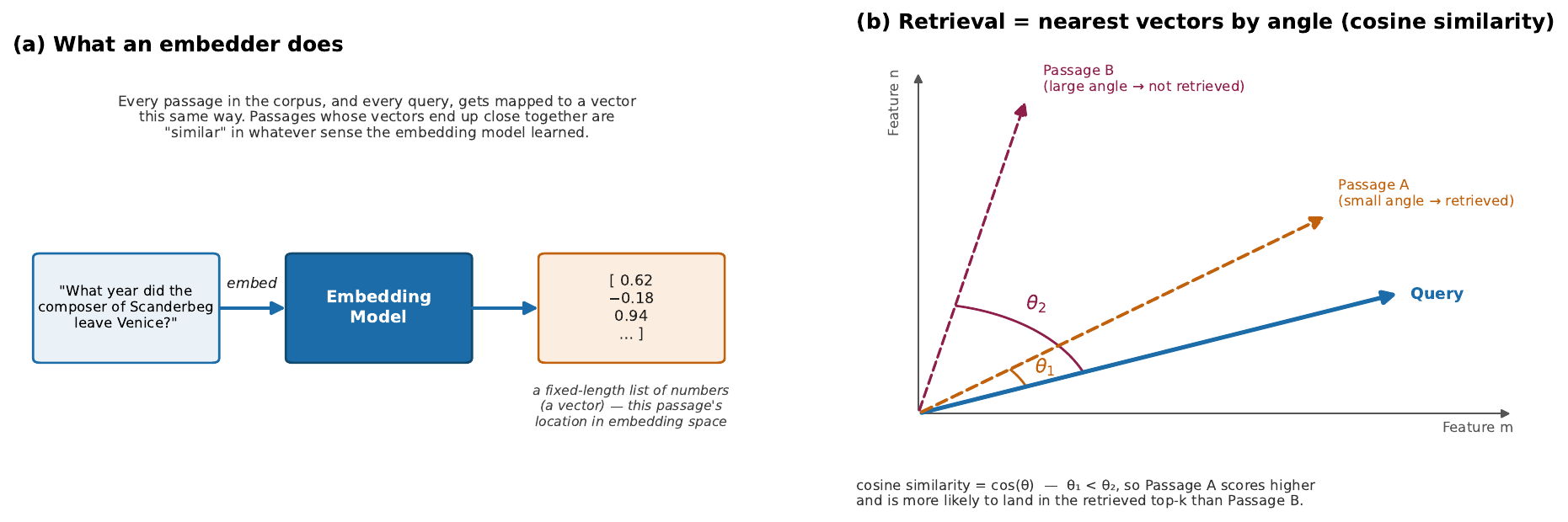}
\caption{What an embedder does (a), and how cosine similarity turns embedded vectors into a ranked retrieval result (b).}
\label{fig:embedding-explainer}
\end{figure*}

\paragraph{Worked example.} The Scanderbeg question of \S\ref{sec:intro} has three gold supporting passages: "Scanderbeg (opera)," "Orlando furioso (Vivaldi, 1714)," and "Rialto Bridge." If all three land in the top-5 retrieved, Recall@5 $= 3/3 = 100\%$; two of three gives $2/3 \approx 66.7\%$. \paragraph{Format sensitivity.} Query-formatting choice is a finding in its own right, and its size varies by model: Qwen3-VL-Embedding-8B swung from 48.3 to 59.9 Recall@5 (11.6 points) purely from query-instruction wording, while BGE-M3 barely moved (54.22 bare $\to$ 54.93 instruction-prefixed; the 54.87 reported in our pre-correction run differs by the $+0.06$ of Table~\ref{tab:reproduction}, and 54.93 is the figure carried everywhere in this paper), consistent with BGE-M3's documentation, which states no instruction prefix is required. Each API provider's asymmetric query/document mode is similarly not optional without a comparable penalty. In asymmetric mode the caller tells the provider whether a given string is a question or a passage (Cohere's \texttt{input\_type}, Gemini's \texttt{task\_type}), and the provider embeds the two roles differently, since a short question and a long document are not the same kind of object. Embedding a question as though it were a document is a silent misuse that costs real Recall. "The embedder's score" is not a single fixed number independent of how a practitioner prompts it.

\section{Realistic Scale: Institutional Data Volumes}
\label{app:tam}

The illustrative corpus sizes of \S\ref{sec:res-index-cost}--\ref{sec:res-embed-cost} (5GB--1TB, 10GB) are round numbers, not claims about any specific institution. Establishing that "institutional scale" is a real category requires a text-only figure, which is harder to come by than it should be. The Library of Congress~\cite{loc2022} reports 21 petabytes of digital collections across 914 million files (2022), but that spans all content types and is dominated by scanned page images, not indexable text, so it cannot anchor a text-corpus extrapolation. Survey evidence for regulated industries (64\% of surveyed organizations manage at least 1 petabyte of total data \cite{informationweek2024}) has the same problem twice over: it is a vendor-sponsored survey, and it counts total data, not text.

We therefore measured one directly, from a public source any reader can re-run. Harvey's Legal Agent Benchmark (LAB)~\cite{harvey2026lab} publishes synthetic corporate-legal matters, each with the document set a firm would receive in a real engagement. Extracting text from every \texttt{.docx}, \texttt{.xlsx}, \texttt{.pptx}, \texttt{.eml} and \texttt{.txt} file in its two largest diligence matters gives:

\begin{table}[H]
\centering
\small
\begin{tabular}{@{}L{2.9cm}rrr@{}}
\toprule
Matter & Files & Extracted text & $\sim$Tokens \\
\midrule
Gaming strategic acquisition & 3,391 & 225.4\,MB & 56.3M \\
Restaurant PE buyout & 3,378 & 217.8\,MB & 54.5M \\
\midrule
Two matters & 6,769 & 443.2\,MB & 110.8M \\
\bottomrule
\end{tabular}
\caption{Extracted text volume of two single legal-diligence matters, measured from the public LAB corpus. Token counts use a 4-characters-per-token approximation (extracted-text sizes in this appendix are decimal MB, the convention its printed token counts imply; reading them as binary shifts every figure by $\approx$5\% and the matched-scale factor's 7.5x lower bound to $\approx$7.2x).}
\label{tab:tam-legal}
\end{table}

The unit that matters is per \emph{matter}: one deal's data room is roughly 225MB of extracted text, forty times the entire 5.64MB MuSiQue corpus this paper's rates are derived from. On that basis the 5GB illustrative corpus of \S\ref{sec:res-index-cost} corresponds to about 23 concurrent matters, and 1TB to roughly 4,650, volumes a mid-sized transactional practice reaches without being an outlier.

\paragraph{The same measurement in dollars.} Megabytes are the wrong unit for the person approving the budget. Applying this paper's own rates (the \$/MB indexing rates of Table~\ref{tab:indexing-cost} and the per-token embedding rates of Table~\ref{tab:embedding-cost}) to one measured matter, and then to the volumes above:

\begin{table}[H]
\centering
\small
\begin{tabular}{@{}L{3.0cm}rr@{}}
\toprule
Volume & Embedding & Indexing \\
\midrule
One matter & \$1--\$8 & \$64--\$997 \\
23 matters ($\approx$5GB) & \$26--\$194 & \$1,460--\$22,900 \\
100 matters/year & \$113--\$845 & \$6,350--\$99,700 \\
4,650 matters (1TB) & \$5.2K--\$39.3K & \$296K--\$4.64M \\
\bottomrule
\end{tabular}
\caption{The same corpus priced two ways, per re-index. Embedding ranges across the four API providers of Table~\ref{tab:embedding-cost} and is \$0 for every self-hostable model; indexing ranges across the six disclosed or reconstructed rates of Table~\ref{tab:indexing-cost}. Like Table~\ref{tab:indexing-cost} these are linear extrapolations, and for graph-construction systems that is an optimistic $O(N)$ assumption: if entity extraction and community detection scale super-linearly with graph density, the true figures are higher, and disproportionately so in the bottom row.}
\label{tab:tam-dollars}
\end{table}

Three things follow that the megabyte figure does not make obvious. First, the two cost lines are not the same size: indexing one matter costs between eight and several hundred times what embedding it costs, so a buyer comparing embedding providers is optimising the smaller line. Second, the indexing range is not a spread between cheap and expensive vendors. It is, in part, the same system priced two ways, since GraphRAG's own two reconstructable estimates differ by 8.5x. This pair is our token-based reconstruction (\$19.49) against KET-RAG's low-cost figure (\$2.30), not the 11x low-versus-high-performance pair quoted elsewhere, because the per-matter arithmetic here needs the two estimates that bracket a single configuration choice (\S\ref{sec:res-index-cost}). At one matter that uncertainty is \$690 and easy to ignore; at a hundred matters a year it is \$69,000, which is a budget line someone has to defend. Third, these figures recur on every re-index, and a live matter is re-indexed as documents arrive.

A note on extrapolation quality, since \S\ref{sec:res-index-cost} rightly cautions that linear scaling is unreliable for graph construction. The per-matter figures here are the least extrapolated in this paper: 225MB is a 40x step from the 5.64MB corpus the rates were measured on, against roughly 185,900x for the 1TB column (binary, consistent with the 1TB = 1,048,576MB convention stated for Table~\ref{tab:indexing-cost}, not the 177,000x a decimal terabyte would give). If super-linear scaling is real, it distorts the bottom row far more than the top one, and in the direction of underestimating. This is a text-only, per-file, reproducible measurement, not a survey statistic, and it is the anchor the illustrative sizes lacked. Two caveats: LAB's matters are synthetic, constructed to resemble real engagements, not sampled from them; and the two measured here are the largest in that corpus, with its median matter 424x smaller (8 files). The distribution is bimodal, so the per-matter figure above is an upper anchor, not an average.

\section{Panel Re-Measurement and Reproduction}
\label{app:reproduction}

Every row of Table~\ref{tab:recall-panel} is reported on a matched \texttt{title\textbackslash ntext} corpus, and every entrant was additionally measured on a body-text-only corpus so the sensitivity of each model to corpus format is visible, not assumed. Table~\ref{tab:reproduction} gives both; $\Delta$ is run-to-run variation on the matched corpus, signed as re-measurement minus original, which is a reproducibility measure and \emph{not} a measure of format sensitivity. The two must not be read as the same column. Twelve entrants reproduce within 0.6 points, confirming they were indexed correctly in both runs; NV-Embed-v2's $\Delta$ required a dedicated replicate: its \emph{original} (v5) index was the mismeasured text-only one, so the corrected \texttt{title\textbackslash ntext} measurement of 69.55, the figure Table~\ref{tab:recall-panel} reports, had been executed once and had no predecessor to reproduce against. The replicate (2026-08-08) re-ran the full sweep end-to-end, all four query instructions on both corpus formats, with the corpus re-embedded from raw text into a fresh cache. Its \texttt{web-search|title+text} arm, the reported configuration, returned 69.55/78.12, identical to the corrected measurement on every one of the 1,000 per-question scores: $\Delta=+0.00$ (both runs' per-question vectors are in the release: \texttt{nvembed\_variant\_sweep\_perq.json}, \texttt{nvembed\_replicate\_run2\_perq.json}). The $+0.24$ headline gap therefore rests on replicated measurements of the \emph{matched-corpus} configuration on both sides ($\Delta=+0.10$ for Nemotron-3-Embed-8B, $+0.00$ for the anchor). \emph{Replicated} here means that same configuration re-executed from scratch, not merely re-swept; the reported 69.55 is the best \texttt{title\textbackslash ntext} variant of four (\S\ref{sec:task-b}), not a text-only figure.

Corpus format, measured properly as the \texttt{title\textbackslash ntext} minus text-only gap, penalizes nine of the ten models we re-embedded in both formats at 95\% confidence, by between 1.46 and 4.73 points (paired bootstrap, $B=10^5$). The tenth, \texttt{text-embedding-3-large}, loses $0.79$ points with a 95\% interval of $[-0.11, +1.71]$: not distinguishable from zero, and barely above that same model's $0.54$ run-to-run drift, so we do not claim a penalty for it. NV-Embed-v2's 2.46 is unremarkable within that spread, sixth-largest of the ten measured, and smaller than the 2.54 of Nemotron-3-Embed-8B, the entrant that tops our panel. The largest penalties belong to \texttt{nv-embedqa-e5-v5} (4.73) and Llama-Nemotron-Embed-1B-v2 (3.92). This is why the corpus format is stated explicitly throughout: it is a first-order experimental variable for the whole panel, not an idiosyncrasy of one model. A corpus-format mismatch applied to any of the thirteen would have depressed it comparably; uniform format is a first-order harness requirement, not a per-model quirk.

\begin{table}[H]
\centering
\footnotesize
\begin{tabular}{@{}L{3.5cm}rrr@{}}
\toprule
Embedder & title+text & $\Delta$ & text-only \\
\midrule
Gemini embedding-001 & 67.24 & $+0.00$ & 65.78 \\
Llama-Nemotron-Embed-1B-v2 & 63.73 & $+0.00$ & 59.81 \\
Nemotron-3-Embed-1B & 64.32 & $-0.00$ & 60.86 \\
mxbai-embed-large-v1 & 55.71 & $+0.01$ & --- \\
Qwen3-VL-Embedding-8B & 59.88 & $-0.02$ & --- \\
BGE-M3 & 54.93 & $+0.06$ & --- \\
Cohere Embed v4 & 60.21 & $+0.08$ & 57.92 \\
\texttt{nv-embedqa-e5-v5} & 57.69 & $+0.09$ & 52.96 \\
Nemotron-3-Embed-8B & 69.79 & $+0.10$ & 67.25 \\
voyage-3.5 & 54.08 & $+0.11$ & 51.17 \\
text-embedding-3-small & 55.38 & $+0.19$ & 53.48 \\
text-embedding-3-large & 59.48 & $-0.54$ & 58.69 \\
\midrule
NV-Embed-v2 & 69.55 & $+0.00$ & 67.09 \\
\bottomrule
\end{tabular}
\caption{Recall@5 on the matched \texttt{title\textbackslash ntext} corpus and on a body-text-only corpus. $\Delta$ is run-to-run variation on the matched corpus. All thirteen entrants reproduce within 0.6 points and were therefore indexed correctly: twelve against their own original run, the anchor against a dedicated replicate, for the reason given in the text; $\Delta$ is signed re-measurement minus original, so \texttt{text-embedding-3-large}'s $-0.54$ means its original run read 60.02 and the re-measurement 59.48, still far nearer its \texttt{title\textbackslash ntext} figure than its 58.69 text-only one, which is the comparison that establishes it was indexed correctly. NV-Embed-v2's original index was the text-only one, so it had no matched-corpus predecessor; its $\Delta$ comes instead from a dedicated replicate run (2026-08-08): the full pipeline re-executed end-to-end with the corpus re-embedded from raw text into a fresh cache, same configuration. The two runs agree on every one of the 1,000 per-question scores, so $\Delta=+0.00$ exactly: deterministic local inference, in line with the other locally-hosted entrants, whose deltas span $-0.02$ to $+0.10$. The $2.46$ separating its two corpus formats is a format penalty and appears below, not in this column. Where a text-only column is blank the model was measured against its already-cached corpus and not re-embedded in both formats. For NV-Embed-v2 both columns report the \texttt{web-search} query instruction, which is also the best of the four swept instructions on the text-only index (67.09 vs 67.07/66.64/64.07), so the row differences the same configuration and the penalty is not inflated by comparing a swept maximum to an unswept point; mixing instructions across the two columns would make the row's own subtraction meaningless.}
\label{tab:reproduction}
\end{table}

\begin{table}[H]
\centering\footnotesize
\setlength{\tabcolsep}{4pt}
\begin{tabular}{@{}L{3.1cm}rr@{}}
\toprule
Embedder & R@10$-$R@5 lift & 95\% CI \\
\midrule
text-embedding-3-large & $+10.67$ & $[+9.48, +11.87]$ \\
voyage-3.5 & $+9.60$ & $[+8.44, +10.79]$ \\
text-embedding-3-small & $+9.40$ & $[+8.29, +10.53]$ \\
Gemini embedding-001 & $+9.11$ & $[+8.03, +10.22]$ \\
Cohere Embed v4 & $+8.88$ & $[+7.81, +9.97]$ \\
Qwen3-VL-Embedding-8B & $+8.65$ & $[+7.56, +9.78]$ \\
mxbai-embed-large-v1 & $+8.64$ & $[+7.55, +9.77]$ \\
NV-Embed-v2 & $+8.58$ & $[+7.52, +9.65]$ \\
nv-embedqa-e5-v5 & $+8.58$ & $[+7.52, +9.65]$ \\
Nemotron-3-Embed-1B & $+8.48$ & $[+7.42, +9.57]$ \\
BGE-M3 & $+7.97$ & $[+6.95, +9.02]$ \\
Nemotron-3-Embed-8B & $+7.75$ & $[+6.75, +8.77]$ \\
Llama-Nemotron-Embed-1B-v2 & $+7.62$ & $[+6.62, +8.65]$ \\
\midrule
\multicolumn{3}{@{}l}{\emph{Format penalty} (title+text $-$ text-only)} \\
\midrule
nv-embedqa-e5-v5 & $+4.73$ & $[+3.34, +6.13]$ \\
Llama-Nemotron-Embed-1B-v2 & $+3.92$ & $[+2.90, +4.96]$ \\
Nemotron-3-Embed-1B & $+3.46$ & $[+2.49, +4.42]$ \\
voyage-3.5 & $+2.92$ & $[+1.93, +3.92]$ \\
Nemotron-3-Embed-8B & $+2.54$ & $[+1.72, +3.38]$ \\
NV-Embed-v2 & $+2.46$ & $[+1.59, +3.35]$ \\
Cohere Embed v4 & $+2.29$ & $[+1.29, +3.31]$ \\
text-embedding-3-small & $+1.89$ & $[+1.01, +2.78]$ \\
Gemini embedding-001 & $+1.46$ & $[+0.46, +2.47]$ \\
text-embedding-3-large$^{\dagger}$ & $+0.79$ & $[-0.11, +1.71]$ \\
\midrule
\multicolumn{3}{@{}l}{\emph{Simultaneous 95\% CIs on the difference vs the top}} \\
\multicolumn{3}{@{}l}{\emph{entrant} (Dunnett-type, max-$t$ crit.\ 2.78)} \\
\midrule
NV-Embed-v2$^{\dagger}$ & $+0.24$ & $[-1.44, +1.92]$ \\
Gemini embedding-001 & $+2.55$ & $[+0.58, +4.52]$ \\
Nemotron-3-Embed-1B & $+5.47$ & $[+3.93, +7.02]$ \\
Llama-Nemotron-Embed-1B-v2 & $+6.06$ & $[+4.27, +7.85]$ \\
Cohere Embed v4 & $+9.58$ & $[+7.59, +11.58]$ \\
Qwen3-VL-Embedding-8B & $+9.91$ & $[+7.87, +11.95]$ \\
text-embedding-3-large & $+10.31$ & $[+8.14, +12.48]$ \\
nv-embedqa-e5-v5 & $+12.10$ & $[+10.05, +14.15]$ \\
mxbai-embed-large-v1 & $+14.08$ & $[+11.79, +16.38]$ \\
text-embedding-3-small & $+14.42$ & $[+12.29, +16.54]$ \\
BGE-M3 & $+14.87$ & $[+12.69, +17.04]$ \\
voyage-3.5 & $+15.71$ & $[+13.50, +17.92]$ \\
\bottomrule
\end{tabular}
\caption{The paired intervals behind every lift and format-penalty claim in \S\ref{sec:res-panel} and this appendix: Recall@10$-$Recall@5 lift for all thirteen entrants (top block) and the corpus-format penalty for the ten measured in both formats (bottom block). Paired bootstrap over the same 1,000 questions, $B=10^5$, seed 42; produced by \texttt{scripts/panel\_intervals.py} and \texttt{scripts/panel\_simultaneous\_ci.py} from the released per-question vectors; the simultaneous block prints all twelve Dunnett-type intervals instead of asserting their verdict. $^{\dagger}$\,interval includes zero: no penalty is claimed for this entrant. Two apparent oddities are verified, not errors: NV-Embed-v2 and \texttt{nv-embedqa-e5-v5} print identical lift rows because their mean lifts are exactly equal (8.575) and these intervals live on a coarse question-mean grid --- the underlying per-question vectors differ on 311 questions, their standard errors differ (0.545 vs 0.541, \texttt{panel\_derived\_stats.json}), and their format-penalty rows differ; the endpoints coincide at full stored precision and not merely at two-decimal display, the grid being coarse enough for two distinct vectors to land on identical percentiles; and two penalties differ by $\pm0.01$ from subtracting Table~\ref{tab:reproduction}'s rounded cells (cell rounding; the artifact carries unrounded values). These are secondary analyses, reported uncorrected (Table~\ref{tab:recall-panel} caption).}
\label{tab:paired-intervals}
\end{table}

The largest deviation among the thirteen is \texttt{text-embedding-3-large} at $-0.54$, still far nearer its \texttt{title\textbackslash ntext} figure than its text-only one; hosted endpoints are not version-pinned and some drift is expected. Four rows sit at $\pm0.00$ (three hosted entrants, plus the anchor's replicate, which is identical on every per-question score); three of those four are exactly zero before rounding, not merely zero as displayed.

Two incidental notes from re-running the hosted models, recorded because a reproducer needs them. First, \texttt{nemotron-3-embed-8b} is not served on the hosted NIM endpoint, consistent with the model card's statement that the 8B tier is self-host-only. Second, model identifiers are specific to their distribution channel: the HuggingFace name \texttt{nemotron-3-embed-1b-bf16} is not a NIM identifier; the served NIM identifier is \texttt{nemotron-3-embed-1b}. A practitioner moving between the HuggingFace and NIM catalogs should look identifiers up per channel and not assume they carry over.

\section{Extended Cost Notes}
\label{app:costnotes}

\paragraph{DGX Spark anchoring.} An 8B embedding model needs roughly 16GB of VRAM and runs on a consumer GPU well under \$1,000; no organization would buy a 128GB unified-memory workstation for embedding alone. We anchor to the DGX Spark (GB10 Grace Blackwell, 128GB unified memory, \$4,699 as of July 2026) because it is the same appliance hosting the heavier LLM workloads of the planned follow-on: the relevant deployment number is "one owned box for the whole retrieval-and-answering pipeline," not an embedding-only figure.

\paragraph{Answering-LLM configuration (\S\ref{sec:res-f1}).} Every answering-cost figure in Table~\ref{tab:answer-cost} comes from \texttt{gpt-4o-mini} called once per question with \texttt{temperature=0.0} and \texttt{max\_tokens=32}: deterministic decoding, one sample, no self-consistency or reranking. Retrieved passages are serialised as \texttt{[\textit{i}] \{title\}: \{text\}} joined by blank lines. The system prompt is: \textit{"You answer questions using only the provided passages. Answer as concisely as possible --- a short phrase or entity, not a full sentence. If the passages do not contain the answer, give your best guess in as few words as possible."} The user prompt is: \textit{"Passages:\textbackslash n\{context\}\textbackslash n\textbackslash nQuestion: \{question\}\textbackslash nAnswer (short phrase only):"}. Scoring takes the maximum SQuAD token-overlap F1 over the gold answer and its aliases. The instruction to guess rather than abstain is deliberate: it keeps F1 a measure of what the retrieved context supports, not of the model's calibration.

\paragraph{GraphRAG conversion imprecision.} The \$19.49--\$38.99 reconstruction (\S\ref{sec:res-audit}) converts HippoRAG-2's Llama-3.3-70B token counts to gpt-4o-mini rates. Two compounding imprecisions: (1) different tokenizers miscount across models; (2) the two models plausibly extract a different \emph{amount} of entity/relation content from the same corpus. Directional estimate only.

\paragraph{KET-RAG's \$33K trace.} KET-RAG's own paper reports \$21 to index a 3.2MB HotpotQA sample (gpt-4o-mini), a \$6.56/MB rate on a different dataset with different chunking, which is why it is excluded from Table~\ref{tab:indexing-cost}'s MuSiQue-corpus rates. Including it anyway would extend the 1TB top of the range from \$4.64M to $\approx$\$6.9M, so the spread this paper quotes remains the conservative reading, not the widest one. Its widely-repeated "\$33K to index a 5GB legal case" is KET-RAG's own linear extrapolation of that figure (\$21 $\times$ 1,600 $\approx$ \$33,600), citing a legal-tech marketing blog only for the "5GB is a realistic case size" premise. It is gpt-4o-mini-based arithmetic, but it is not the independently-sourced estimate it is often repeated as.

\paragraph{gpt-4o-mini rate caveat.} The batch rates used (\$0.075/1M input, \$0.30/1M output; standard \$0.15/\$0.60) are the rates at which our own runs were billed on 2026-07-21 (the meter agreed with the rate card to within 4\%, \S\ref{sec:res-f1}); the model was no longer listed on OpenAI's public pricing page as of that date, so readers re-running it should re-verify.

\paragraph{\texttt{nv-embedqa-mistral-7b-v2}.} This candidate was not served under our API credentials at measurement time. NVIDIA's NGC catalog metadata (Community License + Apache-2.0, no MS MARCO lineage) indicates it was commercial, but it no longer appears in the NIM support matrix, so we treat it as unavailable, not as a credentials issue, and note it here for anyone attempting the same panel.

\balance


\begingroup
\footnotesize
\begin{thebibliography}{99}
\setlength{\itemsep}{2pt}

\bibitem{brehme2025} Brehme, T., Dornauer, M., St\"{o}hle, T., Ehrhart, T., \& Breu, R. (2025). Retrieval-Augmented Generation in Industry: An Interview Study on Use Cases, Requirements, Challenges, and Evaluation. In \textit{Proceedings of the 17th International Conference on Knowledge Discovery and Information Retrieval (KDIR 2025)}, pp.\ 110--122. arXiv:2508.14066.

\bibitem{cohere2026} Cohere. (2026). Embed v4 Model Card --- Amazon Bedrock. AWS Documentation. \url{https://docs.aws.amazon.com/bedrock/latest/userguide/model-card-cohere-embed-v4.html}
\bibitem{aws2026batch} Amazon Web Services. (2026). Supported Regions and models for batch inference. Amazon Bedrock User Guide. \url{https://docs.aws.amazon.com/bedrock/latest/userguide/batch-inference-supported.html} (accessed 2026-08-15).

\bibitem{cc2013} Creative Commons. (2013). Creative Commons Attribution-NonCommercial 4.0 International Public License. \url{https://creativecommons.org/licenses/by-nc/4.0/legalcode}

\bibitem{databricks2024} Databricks. (2024, June 11). Data Intelligence and AI Trends: Top Products, RAG, and More. \textit{2024 State of Data + AI report}. \url{https://www.databricks.com/blog/data-intelligence-and-ai-trends-top-products-rag-and-more}

\bibitem{google2026} Google. (2026). Gemini Embedding --- API Pricing. Gemini API Documentation. \url{https://ai.google.dev/gemini-api/docs/pricing}

\bibitem{gutierrez2025} Guti\'{e}rrez, B.\ J., Shu, Y., Qi, W., Zhou, S., \& Su, Y. (2025). From RAG to Memory: Non-Parametric Continual Learning for Large Language Models. arXiv:2502.14802.

\bibitem{k2view2024} K2View. (2024). GenAI Adoption 2024: The Challenge with Enterprise Data. \url{https://www.k2view.com/genai-adoption-survey/}

\bibitem{lee2024} Lee, C., Roy, R., Xu, M., Raiman, J., Shoeybi, M., Catanzaro, B., \& Ping, W. (2024). NV-Embed: Improved Techniques for Training LLMs as Generalist Embedding Models. arXiv:2405.17428.

\bibitem{loc2022} Library of Congress. (2022). Digital Collections Management FAQ. (21 petabytes / 914 million unique files, official figure.)

\bibitem{informationweek2024} InformationWeek. (2024, July 22). How Much Data Is Too Much for Organizations to Derive Value? (Reporting AvePoint's \textit{AI \& Information Management Report}: 64\% of surveyed organizations manage at least one petabyte. Vendor-sponsored survey; cited here only to illustrate that available figures measure total data rather than indexable text.) \url{https://www.informationweek.com/data-management/how-much-data-is-too-much-for-organizations-to-derive-value-}

\bibitem{longpre2024} Longpre, S., et al. (2024). A Large-Scale Audit of Dataset Licensing and Attribution in AI. arXiv:2310.16787.

\bibitem{nvidia2025discussion} NVIDIA Corporation. (2025, May 9). Comment by \texttt{nada5} (NVIDIA org) in "Please consider changing the license of this model to something which is fully free software." HuggingFace discussion \#40, \texttt{nvidia/NV-Embed-v2}. \url{https://huggingface.co/nvidia/NV-Embed-v2/discussions/40}

\bibitem{openweights2026} NVIDIA, Microsoft, Meta, IBM, Dell Technologies, Hugging Face, The Linux Foundation, et al. (2026, July 24). Open Weights and American AI Leadership. Joint policy letter, 25 signatories. \url{https://images.nvidia.com/pdf/Open-Weights-and-American-AI-Leadership.pdf}

\bibitem{nvidia2024nvembed} NVIDIA Corporation. (2024). NV-Embed-v2 Model Card. HuggingFace. \url{https://huggingface.co/nvidia/NV-Embed-v2}

\bibitem{nvidia2026nemotron8b} NVIDIA Corporation. (2026). Nemotron-3-Embed-8B-BF16 Model Card. HuggingFace. \url{https://huggingface.co/nvidia/Nemotron-3-Embed-8B-BF16} (release date 07/16/2026 and \enquote{ready for commercial use} per the card; accessed 2026-08-15).

\bibitem{nvidia2026nemotron1b} NVIDIA Corporation. (2026). Nemotron-3-Embed-1B-BF16 Model Card. HuggingFace. \url{https://huggingface.co/nvidia/Nemotron-3-Embed-1B-BF16}

\bibitem{nvidia2026e5} NVIDIA Corporation. (2026). nv-embedqa-e5-v5 Model Card. build.nvidia.com. \url{https://build.nvidia.com/nvidia/nv-embedqa-e5-v5/modelcard}

\bibitem{nvidia2026spark} NVIDIA Corporation. (2026). DGX Spark Product Page. \url{https://www.nvidia.com/en-us/products/workstations/dgx-spark/}

\bibitem{openai2026} OpenAI. (2026). text-embedding-3-large / text-embedding-3-small Model Documentation. \url{https://developers.openai.com/api/docs/models/text-embedding-3-large}

\bibitem{wang2025proprag} Wang, J., \& Han, J. (2025). PropRAG: Guiding Retrieval with Beam Search over Proposition Paths. In \textit{Proceedings of EMNLP 2025}, pp.\ 6224--6239. arXiv:2504.18070.

\bibitem{trivedi2022musique} Trivedi, H., Balasubramanian, N., Khot, T., \& Sabharwal, A. (2022). MuSiQue: Multihop Questions via Single-hop Question Composition. \textit{TACL}, 10, 539--554.

\bibitem{voyage2026} Voyage AI. (2026). voyage-3.5 Embedding Model Documentation. \url{https://docs.voyageai.com/docs/embeddings}

\bibitem{wu2026sag} Wu, Y., Li, J., Liang, X., Chen, Y., Liang, Y., Mo, L., \& Li, G. (2026). SAG: SQL-Retrieval Augmented Generation with Query-Time Dynamic Hyperedges. arXiv:2606.15971 (preprint, no accepted venue as of this writing).

\bibitem{anthropic2026embeddings} Anthropic. (2026). Embeddings. \textit{Claude Developer Platform documentation}. \url{https://platform.claude.com/docs/en/build-with-claude/embeddings} (accessed 2026-08-14).

\bibitem{yang2018hotpotqa} Yang, Z., Qi, P., Zhang, S., Bengio, Y., Cohen, W.\ W., Salakhutdinov, R., \& Manning, C.\ D. (2018). HotpotQA: A Dataset for Diverse, Explainable Multi-hop Question Answering. In \textit{Proceedings of EMNLP 2018}, pp.\ 2369--2380. arXiv:1809.09600.

\bibitem{harvey2026lab} Harvey. (2026). Legal Agent Benchmark (LAB). \url{https://github.com/harveyai/harvey-labs}. Public corpus of synthetic corporate-legal matters with expert-written rubrics; the two diligence matters measured in Appendix~\ref{app:tam} are \texttt{diligence/gaming-strategic-acquisition} and \texttt{diligence/restaurant-pe-buyout}.

\bibitem{huang2025ketrag} Huang, Y., Zhang, S., \& Xiao, X. (2025). KET-RAG: A Cost-Efficient Multi-Granular Indexing Framework for Graph-RAG. In \textit{Proceedings of the 31st ACM SIGKDD Conference on Knowledge Discovery and Data Mining (KDD 2025)}, Toronto, Canada, August 3--7, 2025. DOI: 10.1145/3711896.3737012. arXiv:2502.09304.

\bibitem{edge2024graphrag} Edge, D., Trinh, H., Cheng, N., Bradley, J., Chao, A., Mody, A., Truitt, S., Metropolitansky, D., Ness, R.\ O., \& Larson, J. (2024). From Local to Global: A Graph RAG Approach to Query-Focused Summarization. Microsoft Research technical report. arXiv:2404.16130.

\end{thebibliography}
\endgroup
\end{document}